\documentclass[
aip,
rsi,
amsmath,amssymb,
reprint,
floatfix,
groupedaddress,
]{revtex4-1}

\usepackage{graphicx}
\usepackage{dcolumn}
\usepackage{bm}

\usepackage{blindtext}
\usepackage{todonotes}

\usepackage{color}
\def\1#1{{\color{red}#1}}
\def\2#1{{\color{blue}#1}}
\usepackage{tikz}

\newcommand{\ttxt}[1]{\text{\tiny{#1}}}
\newcommand{\infint}[2]{\int\limits_{-\infty}^{+\infty}\!#1\,\text{d}#2}

\newcommand{\dom}[0]{\Delta\omega}

\newcommand{\omegan}[0]{\omega_\ttxt{0}}

\newcommand{\Esh}[0]{E_\ttxt{SH}}

\newcommand{\Eshf}[0]{E_\ttxt{FROG}}

\begin{document}

\preprint{Rev Sci Instrum Draft}


\title[Pulse Retrieval Algorithm for iFROG based on DE]{Pulse Retrieval Algorithm for Interferometric Frequency-Resolved Optical Gating Based on Differential Evolution} 



\author{Janne Hyyti}
\author{Esmerando Escoto}
\author{G\"unter Steinmeyer}
\email[]{steinmey@mbi-berlin.de}
\affiliation{Max Born Institute for Nonlinear Optics and Short Pulse Spectroscopy,
Max-Born-Stra{\ss}e 2a, 12489 Berlin, Germany}


\date{\today}

\begin{abstract}
A novel algorithm for the ultrashort laser pulse characterization method of interferometric frequency-resolved optical gating (iFROG) is presented.
Based on a genetic method, namely differential evolution, the algorithm can exploit all available information of an iFROG measurement to retrieve the complex electric field of a pulse. The retrieval is subjected to a series of numerical tests to prove robustness of the algorithm against experimental artifacts and noise. These tests show that the integrated error-correction mechanisms of the iFROG method can be successfully used to remove the effect from timing errors and spectrally varying efficiency in the detection. 
Moreover, the accuracy and noise resilience of the new algorithm are shown to outperform retrieval based on the generalized projections algorithm, which is widely used as the standard method in FROG retrieval. The differential evolution algorithm is further validated with experimental data, measured with unamplified three-cycle pulses from a mode-locked Ti:sapphire laser. Additionally introducing group delay dispersion in the beam path, the retrieval results show excellent agreement with independent measurements with a commercial pulse measurement device based on spectral phase interferometry for direct electric-field retrieval. Further experimental tests with strongly attenuated pulses indicate resilience of differential evolution based retrieval against massive measurement noise.
\end{abstract}

\pacs{}

\maketitle 

\section{Introduction}

\noindent Frequency-resolved optical gating (FROG)~\cite{trebino_book_2000}, along with the numerous variants thereof, is perhaps the most widespread characterization techniques suitable for ultrashort laser pulses and has been demonstrated with pulses as short as two optical cycles in length~\cite{baltuska_amplitude_1998}. The method features two relatively delayed beams, crossing at a finite angle in a nonlinear crystal. The spectrum of the selected nonlinear interaction is recorded as a function of the delay, producing a two-dimensional spectrogram, or a FROG trace.
Only the cross term of the two incident beams is measured while the fundamental light and the nonlinear signal from each individual beam are ignored.
An iterative algorithm is subsequently employed to reconstruct the complex phase and amplitude of the electric field of the test pulse using the recorded FROG trace. Since the introduction of FROG in the 1990's, extensive efforts have been made to refine the pulse retrieval algorithm of FROG to perfection.

In contrast to the standard noncollinear FROG, the interferometric FROG (iFROG)---employing collinearly propagating beams---has seen relatively little progress in terms of algorithm development since its conception at the turn of the millenium\cite{fittinghoff_collinear_1998, amat-roldan_ultrashort_2004}.
This lack of progress owes to the fact that the standard FROG algorithms cannot be easily applied for the more complex iFROG trace, which, in addition to the cross term, also contains the nonlinear signal from the individual beams, leading to the presence of interference fringes in the measured intensity.
On the other hand, this additional signal content also holds more information, resulting in an increased data redundancy and additional error correction capabilities\cite{stibenz_structures_2006}.
While pulse retrieval remains complicated, the use of collinear beam geometry gives iFROG an edge as it allows for tight focusing, which is helpful for measuring pulses with low peak power or when extremely thin nonlinear crystals need to be employed. The latter is often the case for characterization of few-cycle pulses, which may additionally benefit from the absence of geometrical distortions that cam lead to temporal smearing artifacts. Aside from simple pulse characterization, iFROG has been employed in experiments ranging from the measurement of nanoscale light sources created via surface plasmon polaritons (SPP)---facilitating optical waveform control on the femtosecond and nanometer scales \cite{berweger_femtosecond_2011}, and the measurement of SPP dispersion\cite{schmidt_adiabatic_2012}---to the investigation of noninstantaneous $\chi^{(3)}$ processes\cite{hofmann_noninstantaneous_2015}.
In many iFROG experiments, second-harmonic generation (SHG) has been selected as the nonlinear mixing process\cite{schmidt_adiabatic_2012, metzger_high-power_2011}. The benefit of choosing SHG is that a noncollinear FROG trace can be filtered from the SHG iFROG trace, allowing standard FROG algorithms to be used in pulse retrieval\cite{amat-roldan_ultrashort_2004}. Unfortunately, this leads to a diminished dynamic range of the measured data, reducing the significance of the reconstructed detailed pulse shape.
Moreover, this straightforward approach is not applicable with other nonlinear configurations such as third-harmonic generation (THG), as a standard FROG trace cannot be easily extracted from the resulting traces. A second, more sophisticated approach exploits the fundamental modulation component found in SHG iFROG for pulse retrieval\cite{stibenz_interferometric_2005}. The algorithm therein relies on a modified FROG algorithm, where a steepest-descent minimization strategy is used. Unfortunately, finding the gradient for each iteration cycle is computationally expensive, limiting the convergence speed of the algorithm. As both of the above mentioned approaches to pulse retrieval with iFROG rely on a version of the standard FROG algorithm, they may also suffer from caveats of the generalized projections algorithm routinely used in FROG retrieval. Most prominently, in the presence of noise or for complicated pulse structures, the retrieval process can get stuck in local minima. Such situations are difficult and time-consuming to detect and often offer little hope of escaping and finding the correct pulse shape. Furthermore, both retrieval approaches for iFROG traces neglect much of the information contained within an iFROG trace, only using a single modulation component for pulse retrieval.

In this paper, we present a new approach to the already more than a decade old problem of retrieving a pulse from an iFROG measurement, based on a genetic algorithm known as differential evolution (DE)\cite{storn_differential_1997}. We call the retrieval method differential evolution iFROG (DE-iFROG). The full potential of iFROG traces is harnessed with the use of multiple modulation components simultaneously during pulse reconstruction, resulting in an unprecedented accuracy without the cost of a prohibitively long computation time.

The paper is structured as follows: First, the precursory pulse characterization methods leading to the development of iFROG are discussed. We follow with the structure of the iFROG trace and its subtraces, the latter forming the basis of our retrieval algorithm. The genetic algorithm is explained next. We then proceed to present numerical simulations, where the trace is subjected to massive detection noise and systematic errors that could occur in real world measurements, quantifying the effects on the retrieved pulse. Furthermore, we test the accuracy of a standard FROG algorithm using noisy iFROG traces, and compare the results to DE-iFROG. Finally, an experimental demonstration of the DE-iFROG is given with the characterization of unamplified few-cycle pulses. To further demonstrate the accuracy of our method, additional dispersion is introduced to the test pulse via several dispersive optical elements, and the retrieved spectral phases are compared to the known Sellmeier phases. All the results are compared to reference measurements carried out with a commercial spectral phase interferometry for direct electric-field retrieval (SPIDER) \cite{stibenz_optimizing_2006} device.

\section{Precursors to \MakeLowercase{i}FROG}\label{sec:precursors}

The capacity to generate ultrashort laser pulses for a plethora of applications, realized by the development of mode-locking and q-switching techniques, created a demand for metrological tools quantifying the duration of these pulses. Developed in the 1980's, the optical intensity autocorrelation is one of the earliest techniques to answer this demand, and owing to its simplicity, it is still in broad use today.
An autocorrelation measurement splits an investigated pulse in two replicas, imposes a relative delay upon the two, and guides them into a nonlinear medium where harmonic generation takes place. The upconverted light is measured, the intensity of which depends on the temporal overlap of the two pulse replicas. With some assumptions of the shape of the electric field envelope of the pulse, the pulse duration can be estimated.
Two main variants of intensity autocorrelation exist, based on noncollinear~\cite{sala_correlation_1980} and collinear~\cite{diels_control_1985} beam geometries.
The former is commonly known as background-free intensity autocorrelation (AC) as only the nonlinear cross term is measured, approaching zero for a large temporal separation between the two pulses. The latter variant, known as interferometric intensity autocorrelation (iAC), in contrast, measures the upconverted field of each individual pulse in addition to the cross term, resulting in a background independent of the relative delay between the pulses. The advantage of iAC over the background-free variant is that a larger amount of information is recorded, allowing some sensitivity to the spectral phase. Techniques such as MOSAIC~\cite{hirayama_realtime_2002} and PICASO~\cite{nicholson_noise_2002} use this fact to characterize the chirp of femtosecond pulses. In addition, the collinear geometry allows tight focusing, enhancing measurement sensitivity and allowing the characterization of weak pulses. Unfortunately, due to the necessary assumption of a certain pulse shape, neither interferometric nor background-free autocorrelation are capable of unambiguously characterizing the complex electric field of a pulse as a function of time.

Frequency-resolved optical gating is essentially an extension of the background-free autocorrelation, recording not only the intensity, but also the spectrum of the upconverted pulse to create a 2D spectrogram, i.e., a FROG trace~\cite{trebino_book_2000}. The abundance of information contained in a FROG trace results in a redundancy of data, allowing the reconstruction of both the amplitude and the phase of a pulse with no additional information or assumptions necessary. The resulting inverse problem of reconstructing the pulse characteristics from a FROG trace can be solved using iterative algorithms such as generalized projections (GP)~\cite{delong_pulse_1994}. The redundancy of data also makes the pulse retrieval robust against measurement noise, and several ways to detect measurement errors exist, e.g. through marginal checks and by comparison with a separately measured fundamental spectrum. Recently, it was shown that FROG also has an advantage of detecting the presence of pulse instabilities, which plague other techniques that can only measure the coherent artifact~\cite{rhodes_pulseshape_2013}. Techniques such as SPIDER\cite{stibenz_optimizing_2006} avoid the use of an iterative algorithm through the use of Fourier analysis, but sacrifice the ability to perform error checks and are practically blind to the presence of pulse instabilities.

While a FROG measurement corresponds to a spectrally-resolved background-free autocorrelation trace, spectrally resolving an interferometric autocorrelation trace yields a so-called interferometric FROG trace~\cite{stibenz_interferometric_2005}. As an iAC measurement contains more information than an AC measurement, more information is also contained in an iFROG trace compared to its noncollinear counterpart. In fact, a standard FROG trace is contained within the corresponding iFROG trace~\cite{amat-roldan_ultrashort_2004}. As the spectral phase affects the fringe structure of an iFROG trace, attempts have been made to compute the pulse shape analytically from the trace without resorting to iterative algorithms~\cite{amat-roldan_measurement_2005}. Ambiguities in the phase retrieval, however, impose limits to the reliability of such approaches. Moroever, the advantages of FROG are carried over to iFROG, such as the detection and correction of systematic errors in the measured data. Furthermore, iFROG extends the capabilities of FROG further, given the presence of distinct modulation \emph{subtraces}~\cite{stibenz_structures_2006}. All information required to reconstruct the pulse is contained in more than one of these subtraces, increasing the redundancy in the measured data. The subtraces also contain information about the motion of the translation stage, making it possible to correct for timing jitter in the measurement. Compared to noncollinear FROG, an iFROG measurement device is much simpler to construct and align, while the collinearity of the beam geometry insures that no detrimental geometrical beam smearing can occur. Analogously to interferometric autocorrelation, iFROG also allows a tight focusing of the pulse onto the nonlinear crystal, thus greatly extending the sensitivity of the measurement. The high sensitivity makes the measurement of the third-harmonic possible even for nanojoule pulses, opening a window to explore the physics of higher-order nonlinear processes with weak pulses directly from a Ti:sapphire oscillator, a feat which would be difficult to accomplish with a noncollinear beam geometry.

Aside from the established generalized projections, several innovative algorithms have been suggested for FROG, such as particle swarm optimization~\cite{wu_dispersion_2016} and genetic algorithms~\cite{nicholson_evolving_1999}. While different methods to reconstruct a pulse from an iFROG measurement have been introduced~\cite{amat-roldan_ultrashort_2004,amat-roldan_measurement_2005,stibenz_interferometric_2005}, these techniques have not incorporated all available data contained within an iFROG for a single pulse retrieval, thus offering little advantage over the noncollinear FROG. The novel retrieval method presented here, DE-iFROG, however, can use all data that an iFROG measurement has to offer. In the following section we proceed to discuss the structure of an iFROG measurement and explain how the information contained can be used more efficiently for pulse characterization.

\section{Pulse retrieval}\label{sec:pulse retrieval}

\subsection{iFROG trace structure}

The iFROG trace can be written in the form
\begin{equation} \label{eq:ifrog}
I_\ttxt{iFROG}(\omega,\tau) = \left\lvert \infint{\left[ \mathcal{E}(t) + \mathcal{E}(t-\tau)\right]^2e^{-i\omega t}}{t} \right\rvert^2,
\end{equation}
where $\mathcal{E}(t)=E(t)e^{i\omegan t}$ is the analytic electric field, and $\tau$ is the delay. The pulse envelope is described by $E(t)$, and $\omegan$ is the carrier wavelength. The interferometric trace contains several modulational components, which can be analytically separated~\cite{stibenz_structures_2006}. Excluding the modulation with respect to $\tau$, the three bands are
\begin{align}
B_0 &= 2 \left \lvert \Esh(\dom) \right \rvert^2 + 4 \left \lvert \Eshf(\dom,\tau) \right \rvert^2 \label{eq:b0}\\
B_1 &= 8\left \lvert  \Esh(\dom)  \Eshf^*(\dom,\tau) \right \rvert \times \nonumber \\
& \cos (\phi_\ttxt{SH}(\dom) - \phi_\ttxt{FROG}(\dom,\tau) -\frac{\dom\,\tau}{2}) \label{eq:b1}\\
B_2 &= 2\left \lvert \Esh(\dom) \right \rvert, \label{eq:b2}
\end{align}
where $\dom \equiv \omega - 2\,\omegan$, and the complex-valued functions $\Esh(\dom)$ and $\Eshf(\dom,\tau)$ are defined as
\begin{align}
\Esh(\dom) & \equiv \infint{E^2(t)e^{-i\dom t}}{t} \label{eq:Esh}\\
\Eshf(\dom,\tau) & \equiv \infint{E(t)E(t-\tau)e^{-i\dom t} }{t} \label{eq:Efrog}.
\end{align}
The phase of a given complex field $E_i$ is denoted by $\phi_i$.
The modulation term of $B_n$ is given by $\cos [n\times (\omegan + \frac{\dom}{2})\tau ]$, where $n$ denotes the order of the modulation. The different signal frequencies lead to the separation of the three bands in the Fourier domain, and consequently allows the separation of each band through Fourier filtering, cf. Fig~\ref{fig:unchirped pulse}(b).

The zeroth modulation band, $B_0$, also known as the direct-current (DC) band, contains the noncollinear SHG FROG trace $\Eshf$, along with a delay-independent background, comprised of the second harmonic of each of the two individual pulses. Since the SHG FROG trace is gated, the second-harmonic background can easily be obtained at large delay values. As there are only two terms in Eq.~\ref{eq:b0}, this also allows the SHG FROG trace to be separated from the background, and subsequently used for pulse retrieval via any given method used for noncollinear FROG traces~\cite{amat-roldan_ultrashort_2004}. The removal of the harmonic background, however, results in decreased dynamic range of the measurement data. Moreover, none of the additional information contained in the iFROG trace is employed, and thus the potential gains in robustness and sensitivity of the retrieval are wasted.

The fundamental modulation (FM) band, $B_1$, is temporally-gated and is therefore background-free. The FM trace can be obtained from a measured iFROG trace by means of Fourier filtering~\cite{stibenz_structures_2006}. The FM trace offers an increased sensitivity for the phase of the pulse, which can be advantageous in the retrieval of chirped pulses~\cite{stibenz_interferometric_2005}. Unlike all other types of FROG trace, the FM trace is not constrained to positive values.  Moreover, FM traces are quite intuitive in directly indicating the deviation from an ideal flat spectral phase.

While the second-harmonic modulation (SHM) band, $B_2$, does not enable independent retrieval of the pulse shape, it
can nevertheless have an important function. After computing the Fourier transform of the iFROG trace along the $\tau$ ---the first step to separate the modulational bands from the trace---the SHM band takes the form:
\begin{align} \label{eq:B2 dirac delta}
&\infint{ \left\{ B_2 \times \cos  \left[ (2\omegan + \dom) \tau \right]   \right\} e^{-i\kappa\tau} }{\tau} \nonumber \\
&\propto B_2 \times [\delta(\omega-\kappa) + \delta(\omega+\kappa)]
\end{align}
where $\kappa$ is the angular delay frequency, and $\delta$ is the Dirac delta function. Since $B_2$ is independent of $\tau$, one expects a linear dependence between delay frequency and frequency in either of the SHM modulation bands. In other words, any deviation from a straight line is indicative of a jitter in the delay measurement. Using the information in the SHM band, one can further correct for such delay artifacts, as we will show later.

\subsection{Differential evolution algorithm}

Differential evolution (DE) is an genetic search heuristic that has found widespread application in the synthesis of antenna arrays and inverse scattering problems \cite{rocca_differential_2011}. Genetic search algorithms have been inspired by Darwin's evolution, and they feature population which evolve over time by mutation, cross-over, and selection of the fittest individuals. Compared to more conventional search heuristics previously employed for FROG, genetic algorithms are seeded with a large number of initial random guesses rather than a single one. Choosing a sufficiently large initial population can therefore substantially reduce the problem of local stagnation. At the same time, however, genetic algorithms are expected to be rather slow. In order to allow a fair comparison between DE and GP, we therefore limited the convergence to the same time in the examples discussed below.

\begin{figure}[!htbp]
\fbox{\includegraphics[width=.7\columnwidth]{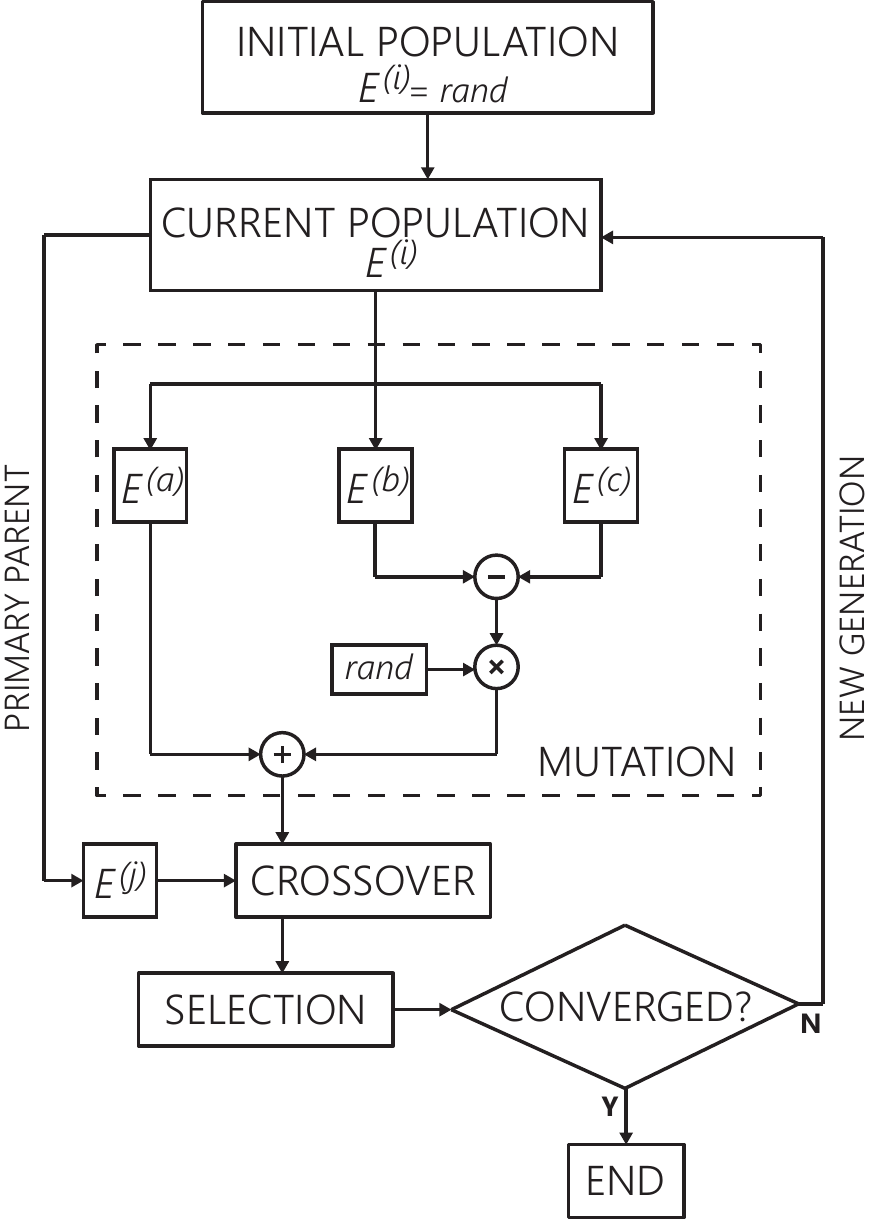}}
\caption{Flowchart of the DE algorithm, with emphasis on the mutation process.}
\label{fig:flow}
\end{figure}

The DE algorithm uses a few simple procedures to refine a population of numeric arrays (individuals) in order to find the fittest individual as close as possible to the global minimum of a given minimization problem. In contrast to other genetic algorithms that typically employ binary digits to encode the genetic material of the population, DE uses floating-point arrays instead, which are more convenient in representing the electric field structure of a pulse. Since its introduction in 1997, DE has spawned countless applications in various fields~\cite{rocca_differential_2011}. The DE algorithm is illustrated by a flowchart in Fig.~\ref{fig:flow}.

In order to adapt DE for iFROG, a population of electric fields $E^{(i)}$ is used, with $i = 1,2, ...\, , D$, where $D$ is the population size. The evolution of the population is guided by measuring the fitness of each individual of the population, allowing fitter individuals to survive while discarding weaker specimen through the process of \textit{selection}. The surviving population (parents) is subsequently used to create new offspring through \textit{crossover} and \textit{mutation} steps. Mutation is performed by randomly selecting three members of the population $E^{(a)}$, $E^{(b)}$, and $E^{(c)}$, arranged by decreasing fitness and used to create a mutant vector $E^{(m)}$:
\begin{equation}
E^{(m)} = E^{(a)} + \rho_m \times \left(E^{(b)}-E^{(c)}\right),
\end{equation}
where $\rho_i\in[0,1]$ is a random number. The difference on the second term of this equation provides variety of solutions by the random number limited by the convergence of the population, hence the name \emph{differential} evolution. This mutant vector is combined with a member of the population to produce an offspring $E_j^{(o)}$ via the crossover step, defined through
\begin{equation}
    E_j^{(o)} =
\begin{cases}
    E_j^{(m)} & \text{if } \rho_o \leq \chi \\
    E_j^{(i)}              & \text{else}
\end{cases}
\end{equation}
where $j$ denotes an element of the array, and $\chi$ is the crossover probability. The fitness of the offspring $E^{(o)}$ is measured, and the offspring are pitted against their parents, resulting to a new generation after the selection step. This evolutionary process preserves qualities of the individuals pertaining to an optimal solution through generations, while weaker qualities are subjected to larger changes with the hope of finding a better replacement, leading to an eventual refinement of the population toward the global minimum, i.e., the electric field of the measured pulse.

In order to measure the fitness of an individual with a given electric field, we compute the root-mean-square errors $G_x$ for the desired subtraces $B_{\rm x}^{\rm (sim)}$ with respect to the corresponding subtraces $B_{\rm x}^{\rm (meas)}$ extracted from the iFROG measurement:
\begin{multline}
G_x = \frac{1}{\sqrt{N M}} \times \\
 \sqrt{\sum_{i=1}^N \sum_{j=1}^M \left[ B_{\rm x}^{\rm{(meas)}}(\dom_i,\tau_j) - \mu B_{\rm x}^{\rm{(sim)}} (\dom_i,\tau_j) \right] ^2 },
 \label{eq:Gval}
\end{multline}
where $\mu$ is the constant that minimizes $G_x$, while M and N are the number of delay and frequency points in the trace, respectively.
Finally, the fitness $G$ is obtained by summing the individual errors:
\begin{equation}
G = \sum_x G_x,
\end{equation}

Each subtrace can be independently computed using Eqs.~\ref{eq:b0}--\ref{eq:Efrog}.
Note that the error can be computed separately for any given set of $\tau$ values.
This brings flexibility to the algorithm, allowing reconstruction from partial traces using only a handful of recorded spectra, similar to recent ptychographic retrieval algorithms for FROG variants~\cite{sidorenko_ptychographic_2016,hyyti_interferometric_2017}. By limiting the number of chosen spectra to a fraction of the necessarily large data set of an iFROG measurement (interferometric precision is required to resolve the fringes of the spectrogram), the computational burden of the retrieval can be significantly reduced. Moreover, should particular regions of a recorded trace have experimental errors or high levels of noise, they can be safely excluded from the retrieval.

The advantage of writing the subtraces in terms of harmonic fields and standard FROG traces is that this results in a greatly reduced computational cost of the fitness measurement. This is mainly due to the fact that the fields $\Eshf$ and $\Esh$ can be computed without explicitly including the rapidly oscillating carrier in the electric field.
The direct implementation of a (genetic) retrieval algorithm, as shown for SHG and PG FROG \citep{nicholson_evolving_1999}, would be highly inefficient and problematic for iFROG. The entire iFROG trace with interferometric precision using Eq.~(\ref{eq:ifrog}) would have to be computed, subsequently employing Fourier-filtering techniques to extract the subtraces, i.e., mimicking the process for the physical measurement. We proceed to give a short example to better explain the computational effort this implies.
The Nyquist delay sampling limit for an iFROG trace, modulated at 1 and 2 times the carrier frequency, is one fourth of an optical cycle. For example, a sampling of $<0.5$~fs is needed for a carrier wavelength of 800~nm, resulting in $N_\tau\approx 370$ delay points for a delay window of 200~fs. For $N_{\omega}=50$ points in the relevant frequency range, a FT with $N_{FT}>300$ points is required for each of the delay points, yielding $>100000$ element matrices with most of the elements holding no relevant information. A total of $N_\tau \times N_\omega \times 2 \times N_\ttxt{bands}$ FTs is required to perform a single fitness measurement, along with other operations, e.g. 70000 FTs for two bands. The use of Eqs.~(\ref{eq:b0}) and (\ref{eq:b1}) reduces the number of FTs to the number of selected delay points $N_\tau$, which can be as few as $<10$, depending on the complexity of the trace. Additionally, two FTs are required to compute the harmonic field $\Esh$. Thus, the computational cost of a fitness measurement is reduced to a fraction using the latter approach.

One further advantage of DE-iFROG is that the length of the electric field is not bound to the delay axis, as is the case with commercial FROG software or the principal components generalized projections algorithm~\cite{kane_simultaneous_1997}. An appropriate combination of the electric field sampling time $\Delta t$ and the number of electric field points $N_E$ ensures that all the features of the pulse envelope are within the time extent $\Delta t \times N_E$ of the retrieval, and that corresponding frequency domain holds all the features of the measured trace, while the number of elements with no relevant data can be reduced to a minimum.

\section{Numerical tests}\label{sec:numerical tests}

\begin{figure}[!htbp]
\fbox{\includegraphics[width=\columnwidth]{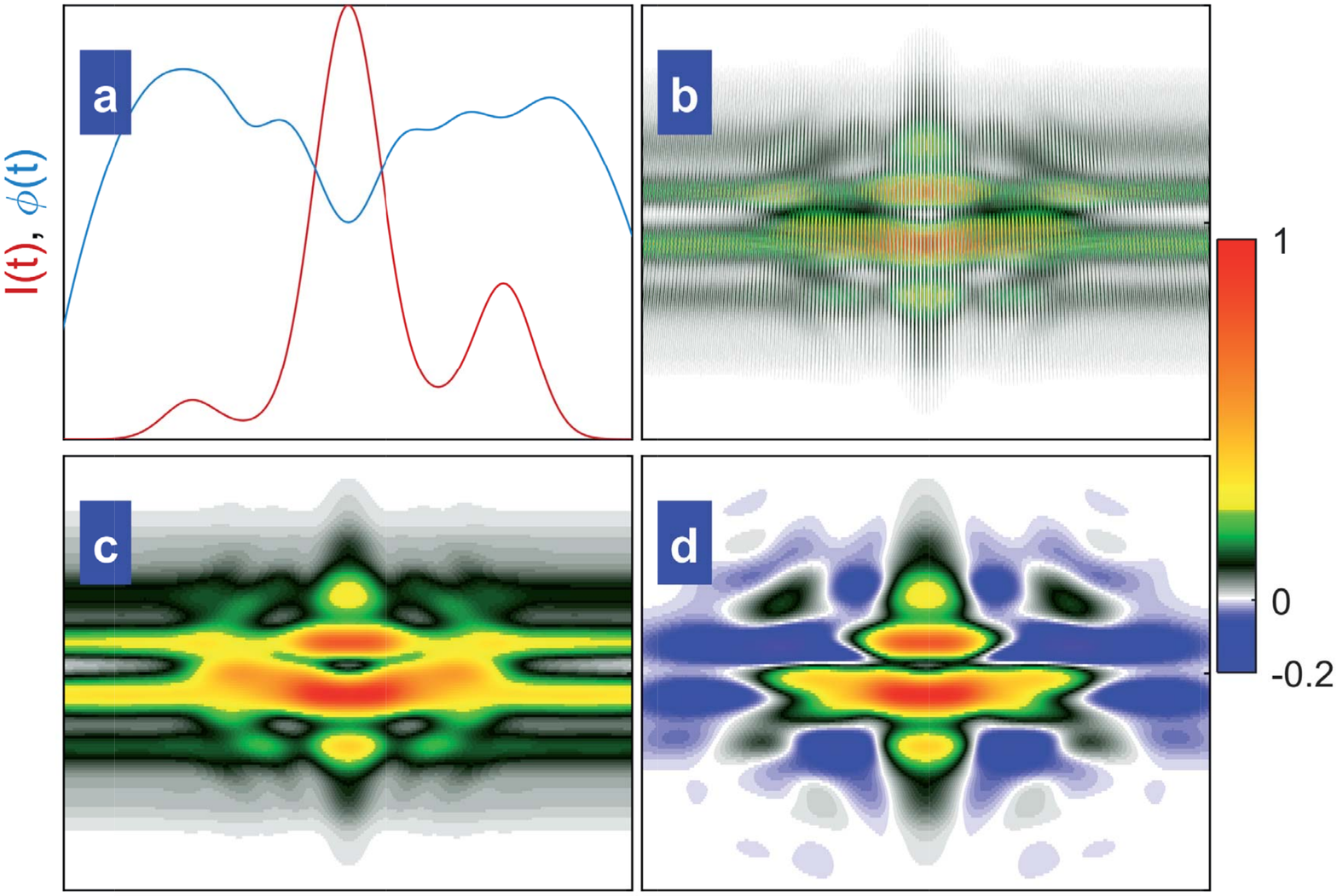}}
\caption{Numerical simulations. \textbf{(a)} The pulse used in the simulations vs. time. \textbf{(b)} The computed iFROG trace for (a). \textbf{(c, d)} The DC (c) and FM (d) subtraces of (b). The traces in (b--d) are plotted for frequency vs. time delay, cf. Fig.~\ref{fig:unchirped pulse}a showing an experimental iFROG trace plotted in the same manner. Time and frequency values are intentionally omitted, as the simulations are applicable for arbitrary pulse lengths and frequency ranges.}
\label{fig:simpulse}
\end{figure}

Numerical simulations were conducted in order to test the robustness of the DE-iFROG retrieval algorithm to noise and other possible error sources that may be present in experimental iFROG measurements. An arbitrary, complex pulse was chosen for the testing, composed of a single strong temporal intensity peak with two weaker satellites on both sides, along with a highly structured phase, cf. Fig.~\ref{fig:simpulse}. The satellites have peak intensities corresponding to 9\% and 36\% of the central peak. Ten DE-retrievals were performed for each set of simulation parameters, and the standard deviations of the retrieved temporal electric field intensities and phases at each point of the time axis was measured.

\begin{figure*}[!htbp]
\fbox{\includegraphics[width=1.2 \columnwidth]{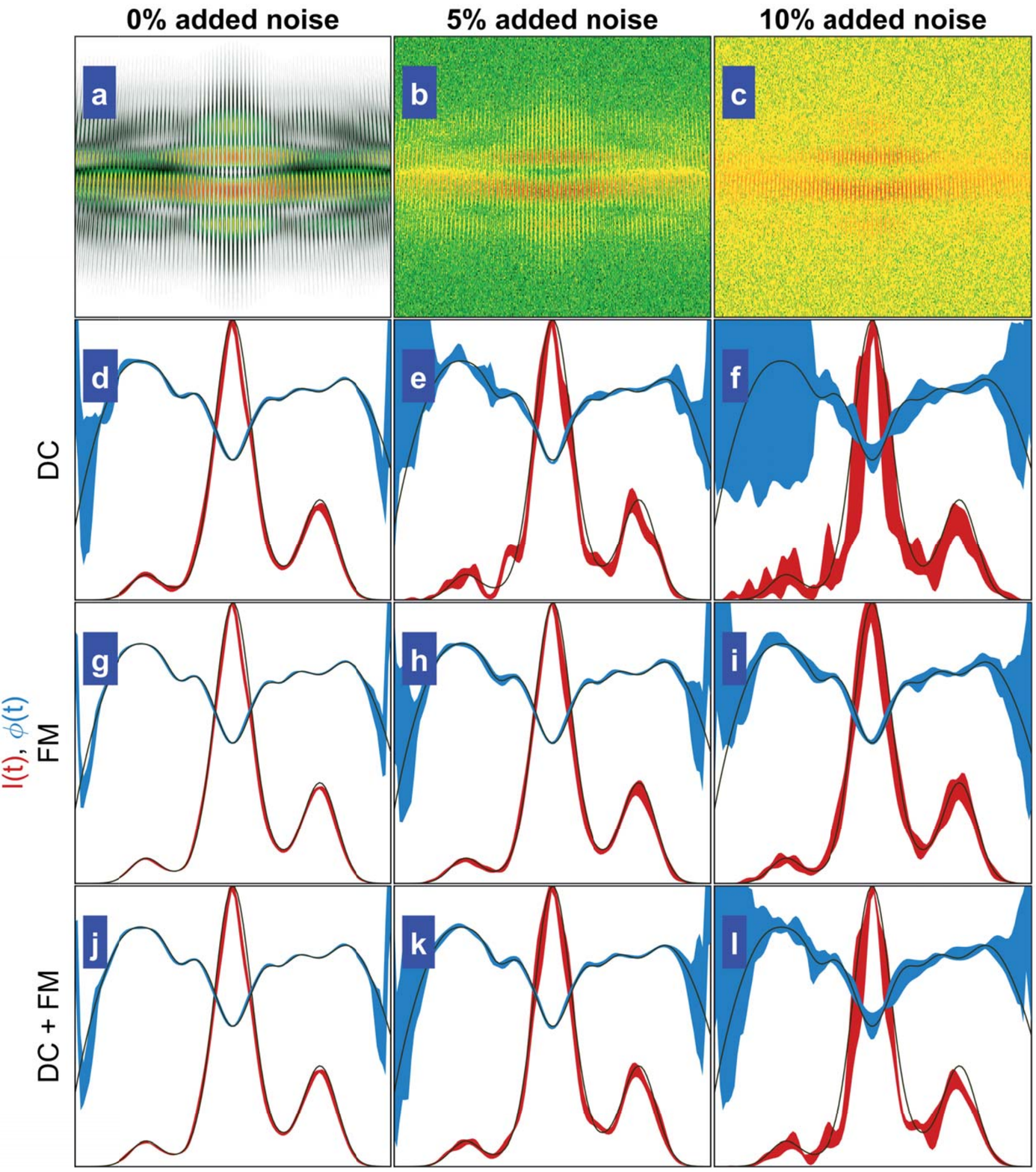}}
\caption{Resilience of DE-iFROG against noise. \textbf{(a)--(c)} Simulated iFROG traces with varying amounts of added noise.
\textbf{(d)--(l)} Retrieved pulses in time domain, with the colored areas corresponding to the intensities (red) and the phases (blue) within one standard deviation of the ensemble. For (d)--(f), only the DC subtrace was used, for (g)--(i) only the FM subtrace, and for (j)--(l) both. The exact intensity and phase of the pulse used to construct the traces are plotted in black.}
\label{fig:noise}
\end{figure*}

\subsection{Noise}

As a first test, normally distributed noise was added to the simulated iFROG trace, see Fig.~\ref{fig:noise}, mimicking noise in the
measurement of the FROG traces. The amount of noise added was controlled by varying the standard deviation of the noise distribution, while keeping the mean at zero. No attempt was made to remove the noise from the simulated traces prior to retrieval with the DE algorithm. This test case is especially interesting, since GP tends to struggle with noise-contaminated FROG traces~\cite{fittinghoff_noise_1995}.
Noise in regions of a FROG trace where no relevant harmonic generation signal should reside can be particularly problematic, resulting in, e.g., high-frequency fluctuations and appearance of ``ghost satellites'' in  the retrieved intensity envelope.
Impressively, DE-iFROG appears to be quite robust against such problems, producing reliable results even with a massive amount of added noise, using a standard deviation of 10\% of the maximum iFROG signal, cf. the rightmost column in Fig.~\ref{fig:noise}.

The noise tolerance of the DE-iFROG was tested using different combinations of the DC and FM subtraces. As mentioned in the previous section, either subtrace can be used by our algorithm. This fact was confirmed by the numerical simulations at hand, as the pulse was retrieved successfully for all test cases with minimal deviations. Moreover, the simulations show that using the modulation subtraces (FM) results in increased robustness against measurement noise, as can be seen by comparing Figs.~\ref{fig:noise}(f) and (i). This robustness can be explained by the visibly more detailed structure of the modulation subtrace. Furthermore, the constant harmonic background in the DC subtrace can obscure features of the underlying standard FROG trace, cf. Eq.~(\ref{eq:b0}). We conclude that the use of modulation subtrace, possibly in combination with the DC subtrace, is beneficial as this results in an increased resilience against measurement noise compared to simply using the DC subtrace. Furthermore, using two subtraces increases the redundancy of the data set, making the retrieval process more reliable.

\begin{figure*}[tbp]
\fbox{\includegraphics[width=1.2\columnwidth]{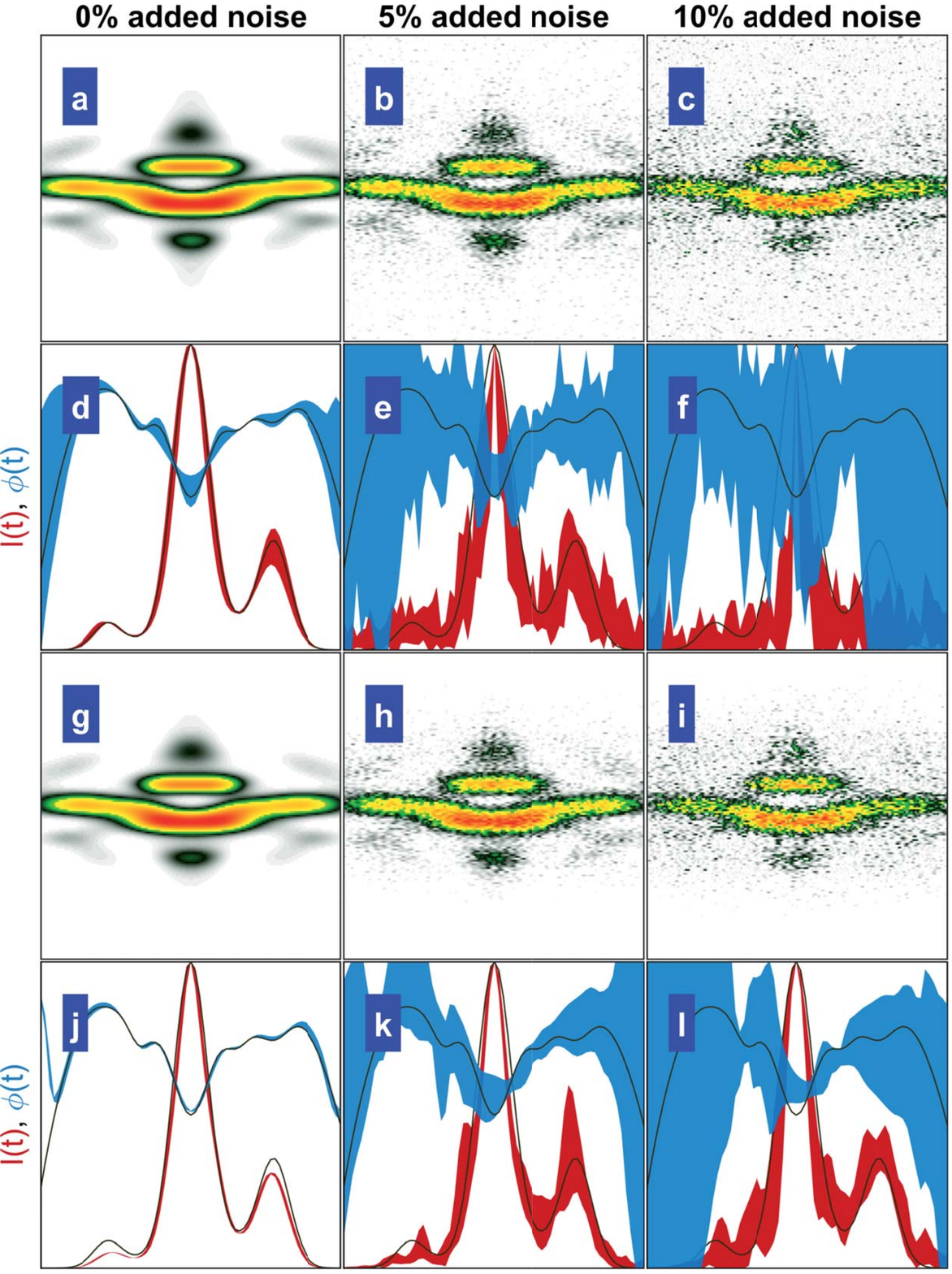}}
\caption{Resilience of GP algorithm against noise. \textbf{(a)--(c)} FROG traces extracted from the iFROG traces in Fig.~\ref{fig:noise}(a)--(c).
\textbf{(d)--(f)} Corresponding retrieved electric fields for (a)--(c). \textbf{(g)--(i)} Same as (a)--(c), but after attempted reduction of noise. \textbf{(j)--(l)} Corresponding retrieved electric fields for (g)--(i).}
\label{fig:gpnoise}
\end{figure*}

In order to further demonstrate the noise resilience of DE-iFROG, we extracted the embedded SHG FROG traces from the same noisy iFROG traces used above (Fig.~\ref{fig:noise}(a)--(c)), and fed these into a standard FROG algorithm based on generalized projections\cite{frog_code}. The retrieval results are shown in Fig.~\ref{fig:gpnoise}. At first, no attempt to remove the noise was made, resulting in the FROG traces in Fig.~\ref{fig:gpnoise}(a)--(c), and the respective retrievals in Fig.~\ref{fig:gpnoise}(d)--(f). Comparing these to the DE-iFROG results in Fig.~\ref{fig:noise}, it is immediately clear that DE-iFROG outperforms the standard FROG extraction method in terms of retrieval accuracy for any given value of added noise, using any combination of subtraces. It is striking that GP based FROG retrieval [Fig.~\ref{fig:gpnoise}(d)] of numerically computed FROG traces without noise is on equal footing as DE-based retrieval from traces with 10\% added noise  [Fig.~\ref{fig:noise}(i)], i.e., often fails to correctly reconstruct the relative intensities of the satellite pulses. Especially for the highest added noise level with a standard deviation of 10\% of maximum iFROG signal, the standard FROG algorithm is quite hopelessly lost with the extracted trace, cf. Fig.~\ref{fig:gpnoise}(f). As a second test, the noise outside the area where actual signal is present was manually removed, and the retrievals repeated, see Fig.~\ref{fig:gpnoise}(g)--(l). This procedure significantly improves the precision of the standard FROG algorithm, but the accuracy is still clearly below DE-iFROG for all noise levels. While the standard deviation of the FROG retrieval in Fig.~\ref{fig:gpnoise}(j) is very low, the satellite peak intensities were underestimated by GP-based retrieval, whereas DE-iFROG retrieves the correct intensity, cf. Fig.~\ref{fig:gpnoise}(d), (g), and (j). It seems that filtering noise, though improving retrieval repeatability, may introduce artifacts that lead to a systematic underestimation of satellite pulses. This analysis shows that DE-iFROG produces much more reliable results for iFROG measurements than the often employed ``extraction of FROG trace $+$ generalized projections'' approach, especially for noisy measurements.

\begin{figure}[tbp]
\fbox{\includegraphics[width=\columnwidth]{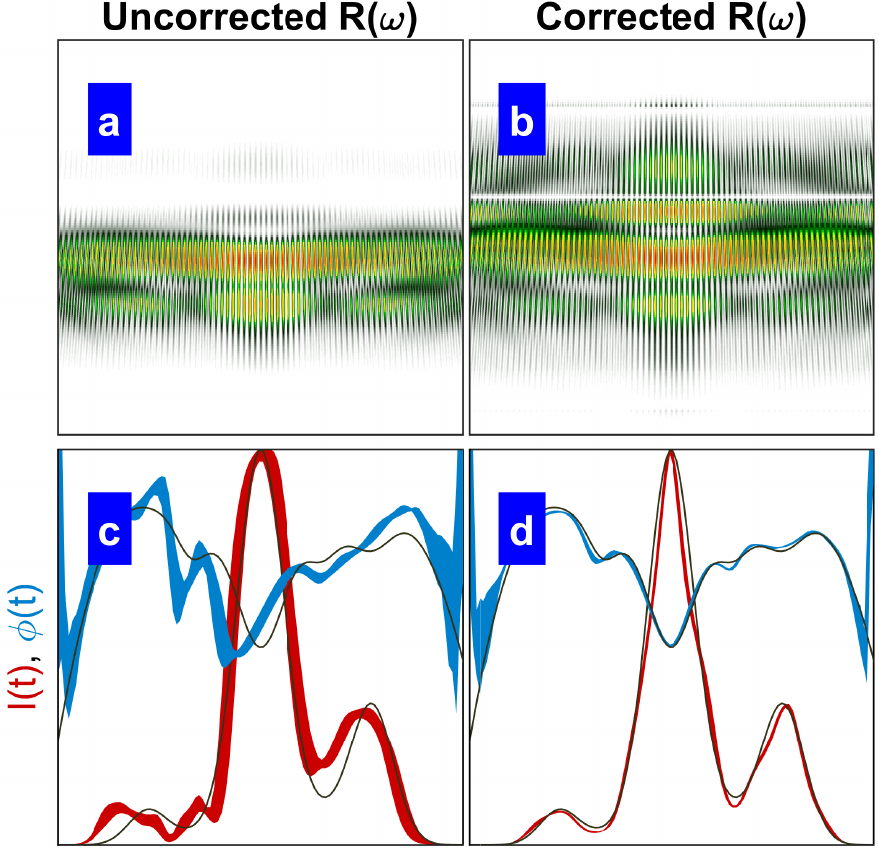}}
\caption{Retrieved pulses from traces with a wavelength dependent scaling \textbf{(a)} without a marginal correction, and \textbf{(b)} with a marginal correction.}
\label{fig:specresponse}
\end{figure}

\subsection{Marginal correction}

A second set of simulations was conducted to estimate the effects of a systematic experimental error in the form of a frequency-dependent transfer function. Such a transfer function is present in virtually any real-world spectral measurement, caused e.g. by the spectral response of a spectrograph, the quantum efficiency of a charge-coupled device typically used to record a spectrum, or distortions of the spectrum as the pulse traverses optical elements. The spectral efficiency of the nonlinear frequency conversion process in a nonlinear crystal is also a major factor, becoming increasingly significant with shorter pulses, especially in the few-cycle regime or when using crystals with limited phase-matching bandwidths. This systematic error can be mitigated in noncollinear FROG traces by comparing the autoconvolution of a separately-measured fundamental spectrum with the frequency marginal of the FROG trace~\cite{trebino_book_2000}. Fortunately, the same can be done for an iFROG trace as the frequency marginal of the noncollinear FROG trace can be extracted. The effect of this correction procedure is illustrated in Fig.~\ref{fig:specresponse}, which shows the retrieved pulses for iFROG traces corrupted by a strong spectral dependency, with and without a marginal correction. To this end, we assumed the characteristic ${\rm sinc}^2$ spectral dependence resulting from an insufficient phase matching bandwidth, which was centered at the central lobe of our pulse structure. Without correcting for this dependence, we see a rather large spread of retrieved pulse shapes and phases [Fig.~\ref{fig:specresponse}(c)], which is reduced by the marginal correction in [Fig.~\ref{fig:specresponse}(d)]. Two horizontal white lines can be seen in the upper part of Fig.~\ref{fig:specresponse}(b), which are not present in the trace for a flat frequency response shown in Fig.~\ref{fig:simpulse}(b). These lines appear because the marginal correction is ill-defined near the zeros of the ${\rm sinc}^2$ function,   making the harmonic signal irrecoverable.
An additional advantage of performing the marginal correction is that for spectral regions where very low or no signal is expected, the noise in a trace will be reduced.

\begin{figure}[tbp]
\fbox{\includegraphics[width=\columnwidth]{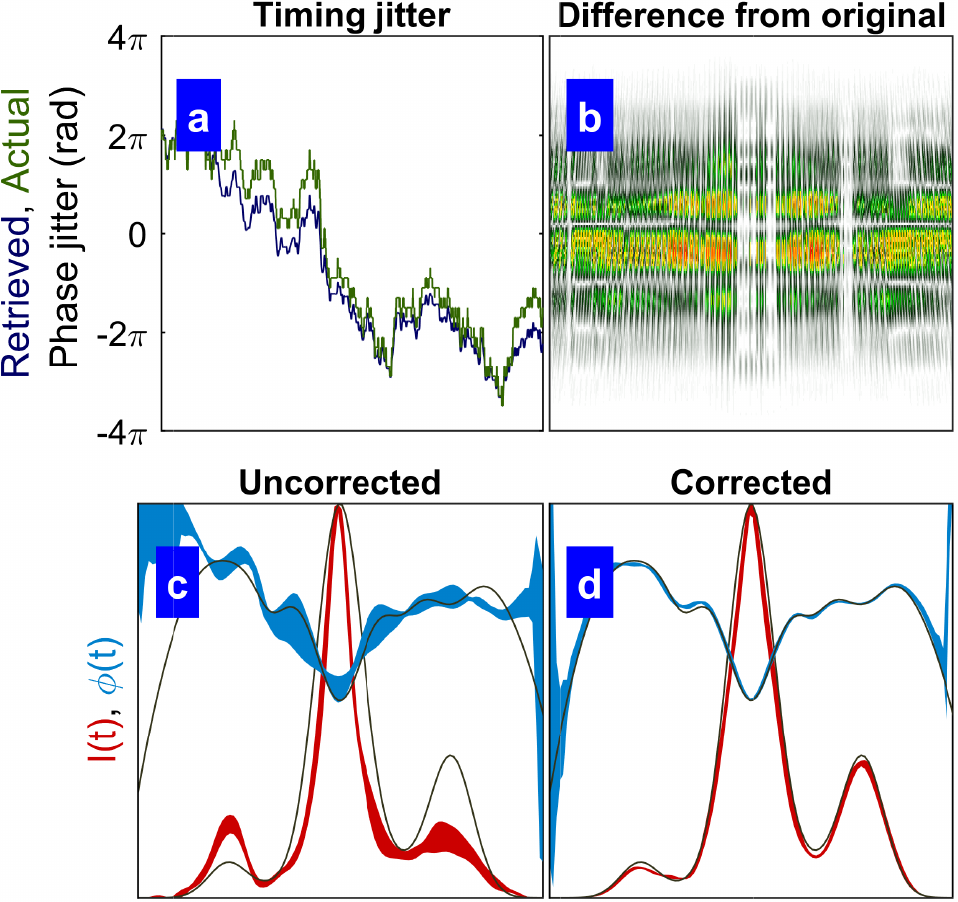}}
\caption{Delay correction. \textbf{(a)} Inserted timing jitter (green line), along with the timing jitter retrieved from the $2\,\omegan$-band (blue line). \textbf{(b)} The intensity difference between the traces with and without timing jitter. \textbf{(c)} Pulse retrieval for the uncorrected trace, and for \textbf{(d)} the trace corrected with the retrieved timing jitter in (a). If the central wavelength is at 800 nm, a 2$\pi$-jitter corresponds to 2800 as. This is a fairly massive amount, as piezoelectric actuators typically have resolutions below 100 as.}
\label{fig:jitter}
 \end{figure}

\subsection{Timing-jitter correction}

Retrieval algorithms for FROG typically assume that the input trace has a perfect delay sampling with a constant delay step. In reality, there will always be an uncertainty in the optical path lengths of the two beams found in any FROG measurement, translating into an uncertainty of the delay between the two pulses. This timing jitter, e.g., caused by mechanical vibrations, fluctuations of air density, and nonuniform movement of a delay stage, distorts each delay step of a FROG measurement. Fortunately, the timing jitter for a given iFROG measurement can be measured and corrected for through Fourier analysis. This correction procedure can be realized by measuring the phase $\phi_\ttxt{mod}= (2\omegan + \dom)\tau$ of the SHM band, i.e. $\cos(\phi_\ttxt{mod})\times B_2$, using the Takeda algorithm~\cite{takeda_fourier-transform_1982}. Multiplying the FM band, $\cos(\phi_\ttxt{mod}/2)\times B_1$, with $\cos(\phi_\ttxt{mod}/2)$ followed by a Fourier filtering step gives $B_1$, thus removing the modulation. Assuming that the spectrometer has been properly wavelength calibrated, the timing jitter can be computed from $\phi_\ttxt{mod}$. This capability of DE-iFROG to measure and correct for a timing jitter is illustrated in Fig.~\ref{fig:jitter}, where a massive jitter of an entire optical cycle is simulated. Impressively, the pulse is perfectly reconstructed after the timing-jitter correction is applied, see Fig.~\ref{fig:jitter}(d).

\section{Experimental measurements}\label{sec:measurements}

\begin{figure}[!htbp]
\fbox{\includegraphics[width=.85\columnwidth]{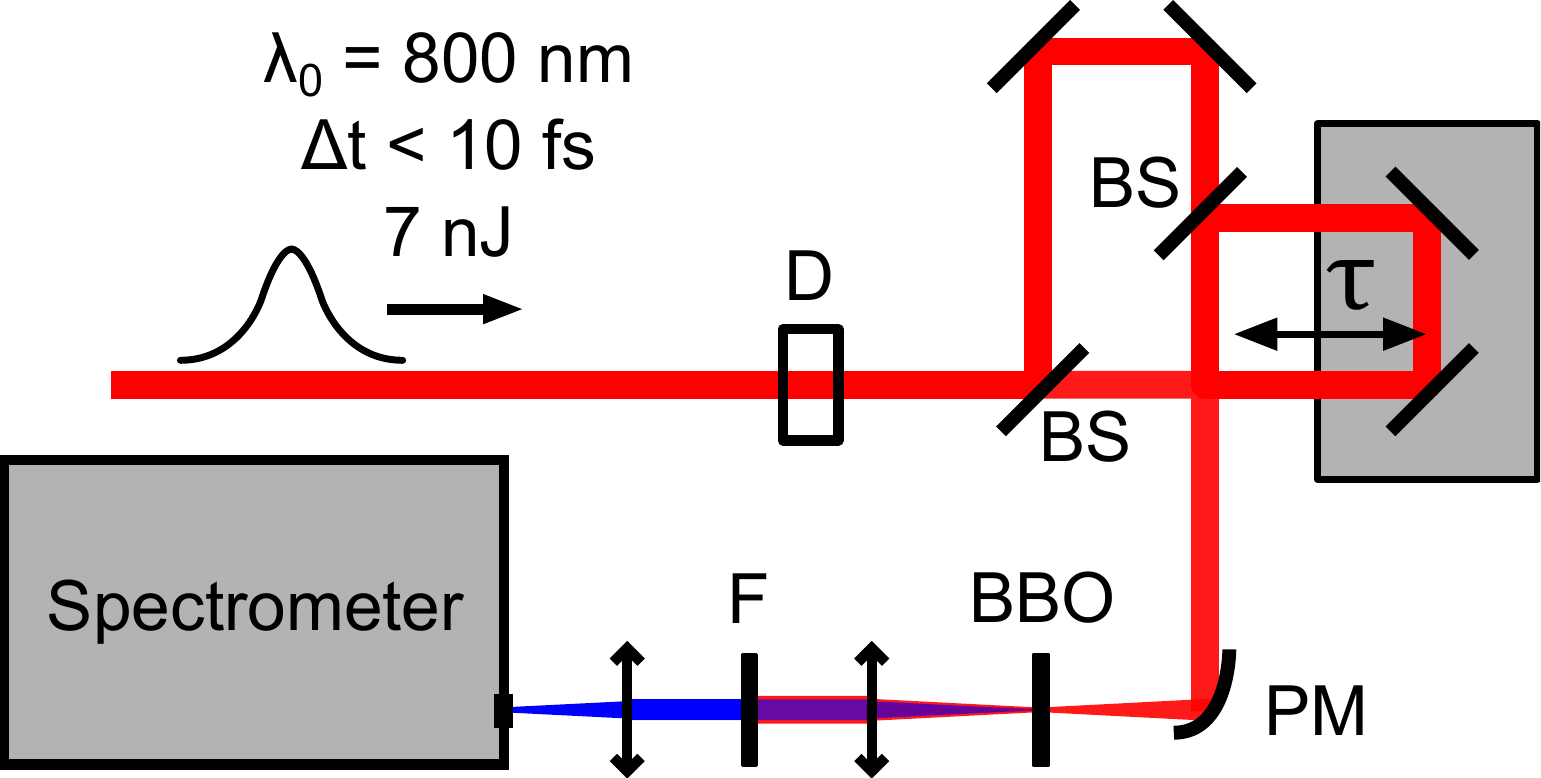}}
\caption{Experimental iFROG setup. Few-cycle pulses from a Ti:sapphire oscillator are split into two replicas with a relative delay via a dispersion-balanced interferometer comprised of a pair of beam splitters (BS) and a piezoelectric delay stage. The collinearly-propagating pulses are focused into a BBO crystal with a parabolic mirror (PM).
After collimation with a lens, spectral and polarizing filters (F) are used to isolate the upconverted light, which is then guided into a spectrometer. For the chirped-pulse measurements, a dispersive element (D) with a known dispersion profile is inserted into the beam path before the interferometer.}
\label{fig:setup}
\end{figure}

To further demonstrate the capabilities of DE-iFROG, experimental measurements were conducted in the challenging domain of few-cycle pulses. A Kerr-lens mode-locked Ti:Sapphire oscillator is used to generate the pulses, the spectra of which are centered at 800 nm. The nearly Fourier-transform-limited pulses are approximately only three optical cycles long, with an energy of 7~nJ. The experimental setup is described in detail in Fig.~\ref{fig:setup}. Separate measurements were also made using a commercial few-cycle SPIDER device (\textit{FC Spider} from APE Angewandte Physik \& Elektronik GmbH). Because of the very broad spectral bandwidth of the pulses, even the dispersion due to air contributes significantly to the broadening of the pulses. The two separate measurements are therefore carefully arranged in order to minimize the optical path difference. To measure the SPIDER interferogram, one of the interferometer arms of the iFROG setup is blocked, and the beam is diverted into the SPIDER device, cf. Fig.~\ref{fig:setup}. The additional dispersion due to air along the path leading to the SPIDER apparatus is taken into account in all the presented results by subtracting this additional dispersion from the spectral phase retrieved by SPIDER.

\subsection{Transform-limited pulse}

\begin{figure*}[tbp]
\fbox{\includegraphics[width=1.2 \columnwidth]{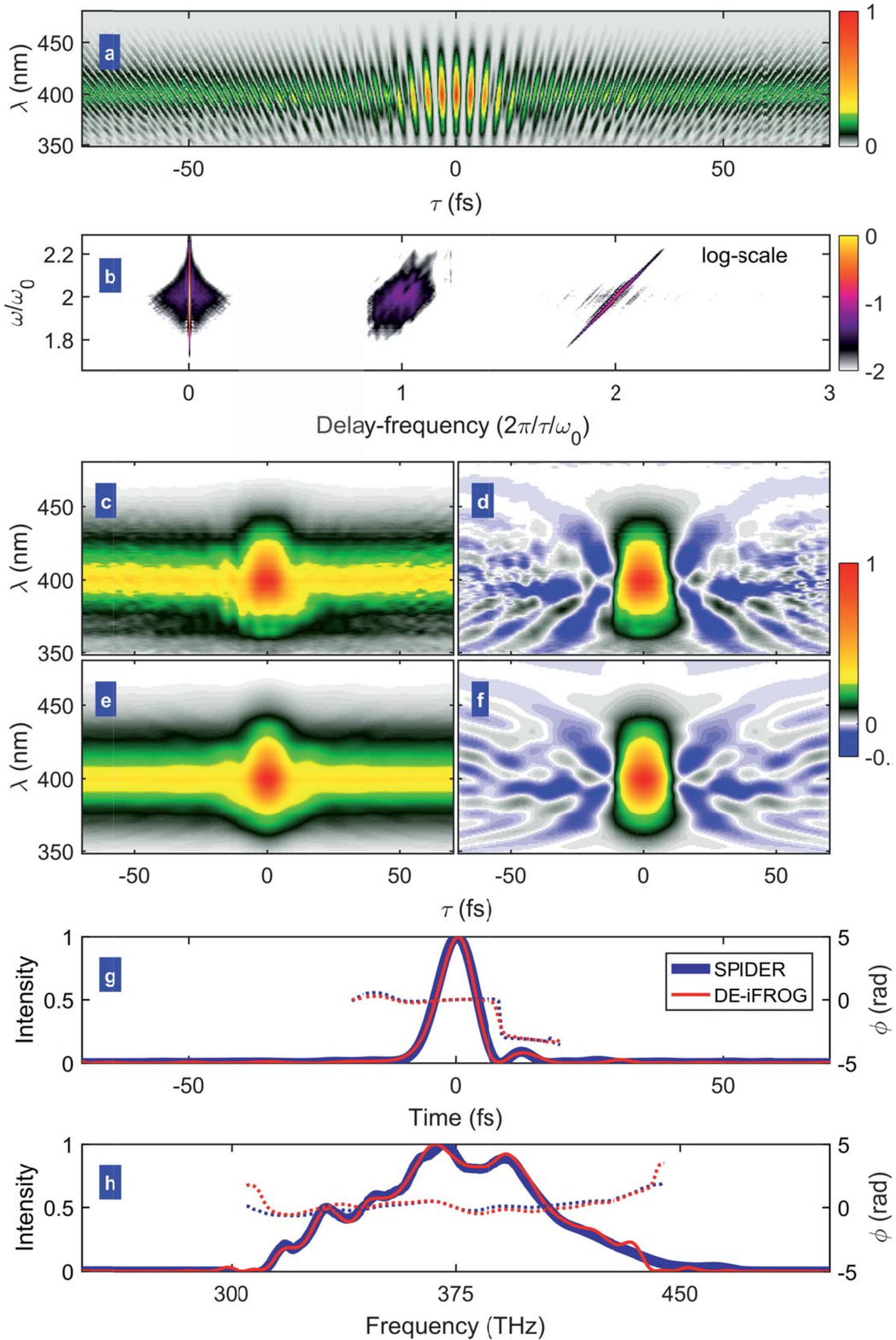}}
\caption{iFROG measurements for a Fourier-transform-limited pulse.
\textbf{(a)} Measured iFROG trace, linear scale.
\textbf{(b)} FT of (a) with respect to $\tau$, logarithmic scale.
\textbf{(c)} Fourier-filtered DC trace using the band at zero delay frequency in (b).
\textbf{(d)} FM subtrace derived from the band at $\omega_0$ delay frequency in (b).
\textbf{(e), (f)} Retrieved DC and FM subtraces, respectively.
\textbf{(g)} Retrieved time-domain electric field intensity (solid lines) and phase (dotted lines) for DE-iFROG (red) and SPIDER (blue).
\textbf{(h)} The same as above, but in frequency-domain. Note that the SPIDER spectral intensity is simply a measured fundamental spectrum after calibration, not a retrieved spectrum. DE-iFROG, on the other hand, retrieves both the phase and the intensity.}
\label{fig:unchirped pulse}
\end{figure*}

The measured iFROG trace for the near Fourier-transform-limited pulse is shown in Fig.~\ref{fig:unchirped pulse}(a), and a Fourier transform of the trace along the delay axis in Fig.~\ref{fig:unchirped pulse}(b). Visual inspection of the Fourier domain tells us that the SHM band (around the delay frequency $2\,\omegan$) is approximately a straight line, suggesting that the delay stage movement is sufficiently accurate. The timing-jitter correction procedure can of course be made to further improve the measured data, as described in the previous section. Each of the two other modulation bands, located at the delay frequencies 0 and $\omegan$, contain the necessary information to reconstruct the complex pulse envelope. Consequently, both the FM and DC subtraces extracted from these bands are employed by the algorithm. The measured and retrieved subtraces, Fig.~\ref{fig:unchirped pulse}(c)--(f), share all the prominent features, but some of the finer details of the measured data are not found in the retrieved subtraces. A closer inspection reveals that most of these details are only found on either positive or negative delays. As an ideal iFROG trace is symmetric with respect to delay, the details found on only a single side of the measured subtraces are therefore only artifacts of the measurement. Furthermore, as the algorithm produces any feature on both sides of the subtraces, any produced value that lies between the two measured values on either side will give the same error for the measurement point. Thus, DE-iFROG will be largely unaffected by one-sided artefacts. Despite the minor measurement artifacts, the $7.6$-fs pulse retrieved by DE-iFROG is in excellent agreement with the SPIDER measurement, see Fig.~\ref{fig:unchirped pulse}(g) and (h).

\subsection{Chirped pulses}

\begin{figure*}[tbp]
\fbox{\includegraphics[width=1.2 \columnwidth]{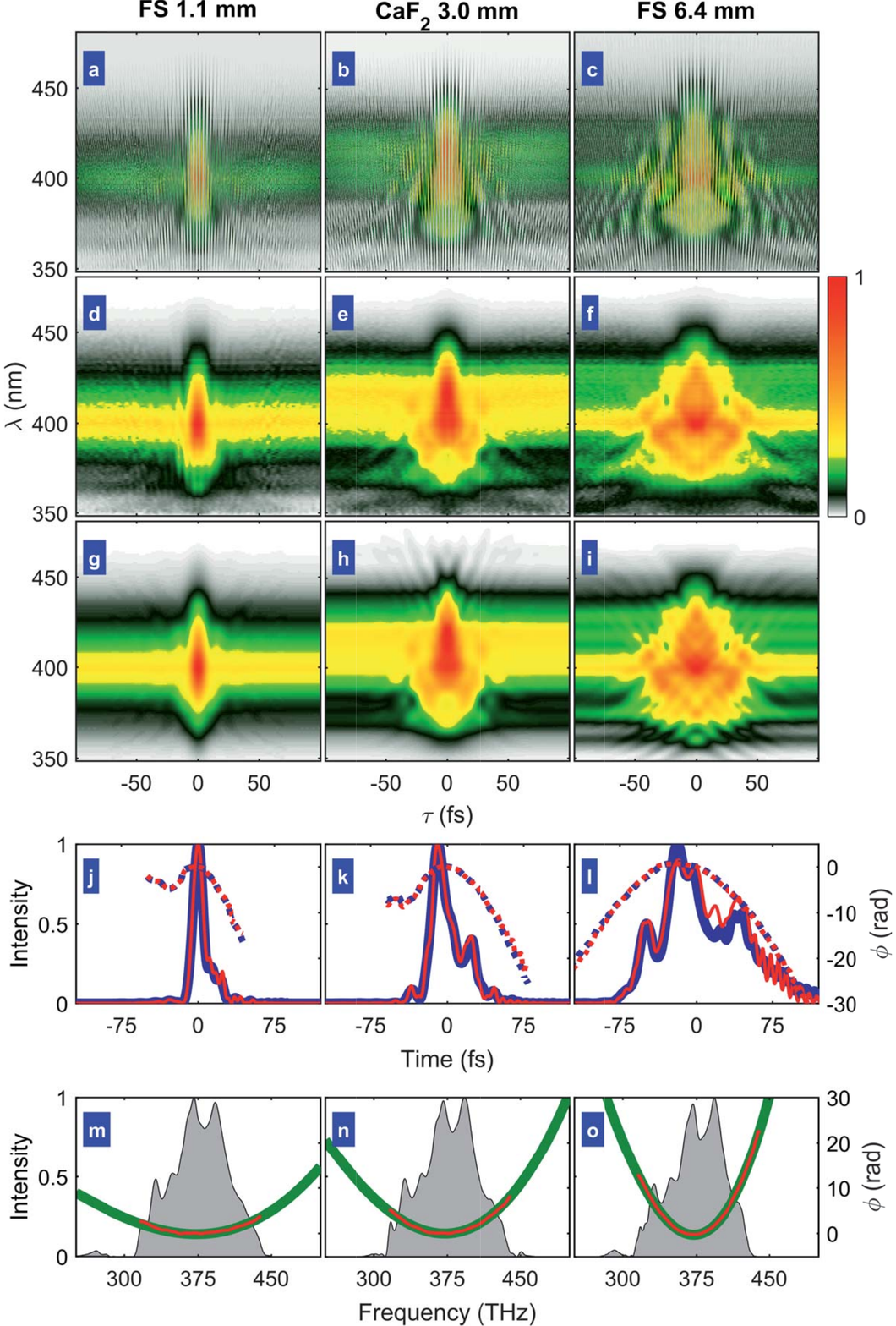}}
\caption{iFROG measurements for chirped pulses. The plots are organized into three columns with each column pertaining to a single measurement employing one of the three dispersive elements, marked on the top of the column. Moving from left to right, the amount of additional dispersion is increased.
\textbf{(a)--(c)} Measured iFROG traces.
\textbf{(d)--(f)} Fourier-filtered DC traces for the traces (a)--(c).
\textbf{(g)--(i)} Retrieved DC traces.
\textbf{(j)--(l)} Retrieved electric field intensities (solid lines) and phases (dotted lines) for DE-iFROG (red) and SPIDER (blue) in time-domain. Note that our SPIDER apparatus is not specified to function with pulses longer than 60~fs, i.e., the SPIDER measurement for the highest dispersion in (o) is subject to error. Nevertheless, good agreement is found for all three cases.
\textbf{(m)--(o)} Retrieved spectral phases after subtraction of the unchirped phase in Fig.~\ref{fig:unchirped pulse}~(h) (red solid lines), and the retrieved spectra (gray). The retrieved phases perfectly match the Sellmeier phases (green solid lines), computed for the respective dispersive elements.}
\label{fig:chirped pulses2}
\end{figure*}

As further test for the DE-iFROG, three different dispersive elements were inserted in sequence into the beam path to introduce a varying amount of group delay dispersion. Two windows of fused silica (FS) with thicknesses $1.1$~mm and $6.4$~mm were used, along with a  $3.0$~mm thick calcium fluoride (CaF$_\mathrm{2}$) window. The three measured iFROG traces for the chirped pulses are shown in Figs.~\ref{fig:chirped pulses2}(a)--(c). Both the DC and FM subtraces were used for the retrieval. The measured and the retrieved DC subtraces are shown in Figs.~\ref{fig:chirped pulses2}(d)--(i). Once again, the DE-iFROG algorithm retrieved artifact-free subtraces while preserving all the significant structures. Consistent with the previous result, the retrieved pulses generally agree well with the SPIDER measurements, cf. Figs.~\ref{fig:chirped pulses2}(j,k). However, for the highest amount of additional dispersion (rightmost column with FS $6.4$~mm) the trailing structure at positive delay shows marked discrepancies between both methods, cf. Fig.~\ref{fig:chirped pulses2}(l). While increasing pulse duration should bring no conceivable problems for DE-iFROG, the commercial SPIDER apparatus is not specified to work for pulses longer than 60 fs as it is specifically designed for few-cycle pulses. Since the retrieved pulse duration exceeds $100$~fs, the rather complex pulses exhibit a time-bandwidth limit of ~15, which seems to exceed the safe limits of the SPIDER apparatus. Nevertheless, phases retrieved by both DE-iFROG and SPIDER still agree well with independent Sellmeier calculations. The resulting dispersion curves are plotted in Fig.~\ref{fig:chirped pulses2}(m)--(o), along with the added dispersion measured by DE-iFROG. The latter values were obtained by subtracting the measured spectral phase of the Fourier-transform-limited pulse (cf. Fig.~\ref{fig:unchirped pulse}(h)) from the chirped-pulse measurements. It is immediately clear that the measured additional dispersion matches almost perfectly with the analytically obtained values.

\subsection{Low signal-to-noise ratio}

\begin{figure}[!htbp]
\fbox{\includegraphics[width=\columnwidth]{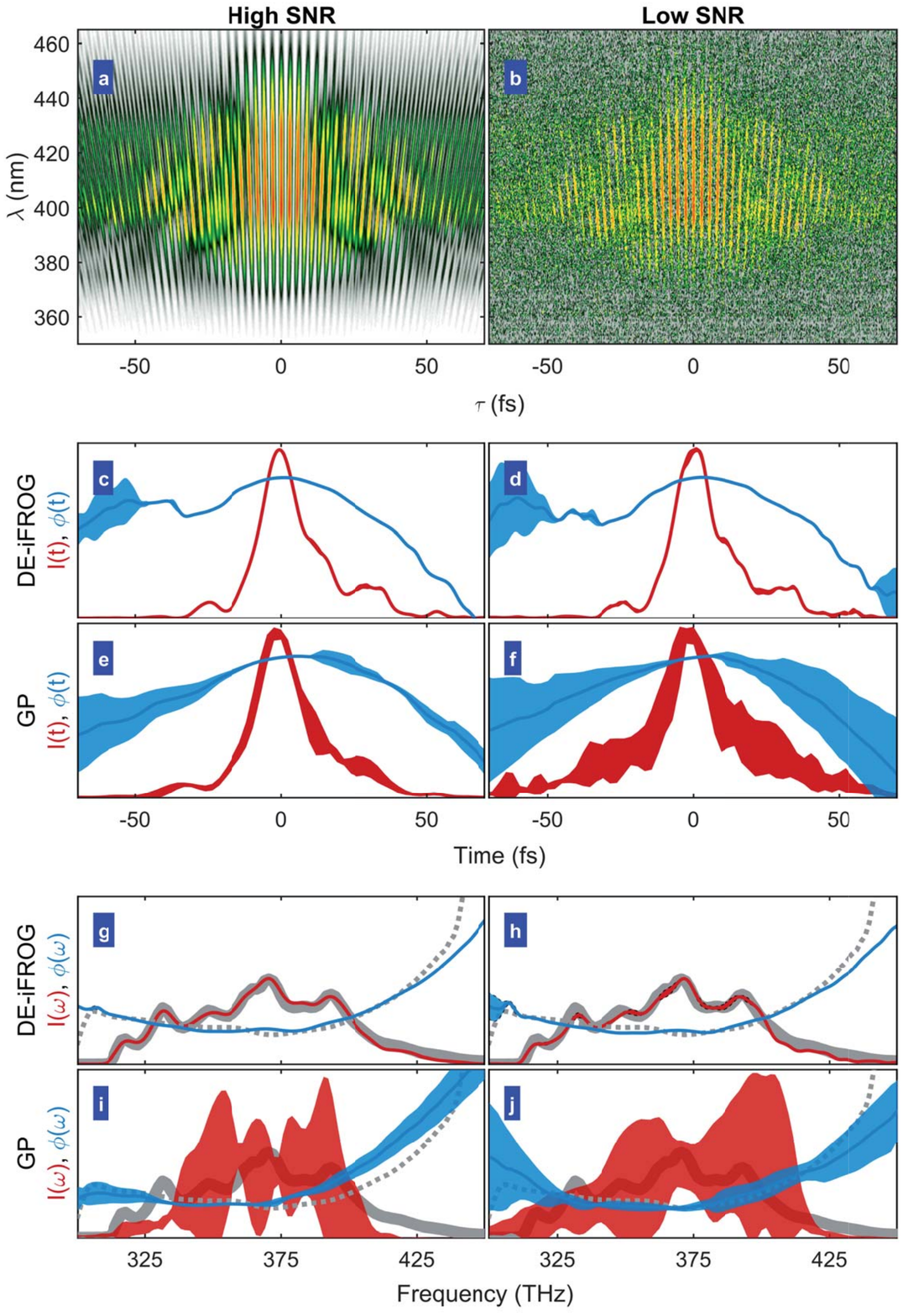}}
\caption{Experiments with high (left column) and low signal-to-noise ratio (right column). \textbf{(a), (b)} Measured iFROG traces with (a) typical system performance, and (b) signal weakened via neutral density filters.
\textbf{(c), (d)} DE-iFROG retrieval results for (a) and (b), plotted as standard deviation versus time.
\textbf{(e), (f)} Same as above, but using GP for the extracted SHG FROG trace.
\textbf{(g)--(j)} Retrieval results in frequency domain. Reference SPIDER measurement of spectral intensity (solid line) and phase (dotted line) using unattenuated pulses is plotted in gray.}
\label{fig:noise experiment}
\end{figure}

Two further iFROG measurements were conducted with strongly varying signal-to-noise ratios (SNR). Here the objective was an experimental confirmation of the numerical simulations presented earlier, indicating that DE-iFROG is much less affected by noise in comparison to generalized projections retrievals. The measurement with high SNR illustrated in Fig.~\ref{fig:noise experiment}(a) is comparable to the previous measurements presented above. The second measurement in Fig.~\ref{fig:noise experiment}(b) has a very low SNR, achieved by deliberately attenuating the iFROG signal with neutral density filters before detection. The standard deviation of the experimental noise relative to the peak signal is approximately $0.1$\% for the high SNR trace whereas the other trace is corrupted by a massive 8\% noise content.

Ten retrievals were made for each iFROG trace and algorithm pair. Only the FM subtrace was used for DE-iFROG, as the numerical simulations indicated this to be the most noise-resistant choice for a subtrace combination. The extracted SHG FROG traces were fed to GP. The obtained statistics yield the standard deviations presented in Fig.~\ref{fig:noise experiment}(c)--(j) for the electric fields in time and frequency domains. Consistent with the numerical simulations, DE-iFROG retrieved very similar pulses for both traces. The inherent ability of DE-iFROG to escape local minima becomes apparent here, as the same optimal solution is reached every time, indicated by the exceedingly low standard deviations of the phase and the intensity in both time and frequency domains. In fact, the standard deviations even for the low SNR retrievals are so low that the means had to be plotted with a thicker line width so that the profiles are visible in the figures.
The slight increases in the standard deviations for the low SNR trace are mostly found for the phases corresponding to very low intensities. GP, on the contrary, suffers from a higher uncertainty in the retrieved time-domain pulse envelopes even for the high SNR measurement, cf. Fig.~\ref{fig:noise experiment}(e) and (f). Moving to lower SNR introduces considerable fluctuations, completely obscuring the underlying true pulse shape.

The differences in the retrieved fields of the two algorithms are most pronounced in the frequency domain. In all cases, DE-iFROG retrieved a spectral intensity and phase that matches the SPIDER measurements almost perfectly, for both high- and low-SNR traces, cf. Fig.~\ref{fig:noise experiment}(g) and (h). The small differences of the retrieved spectral phases are most likely due to unaccounted influences by optical elements guiding the beam to the SPIDER apparatus, and the elements within device itself. This unprecedented agreement between the two methods is especially impressive as no additional information of the spectrum was provided for the DE-iFROG algorithm, in the form of a marginal correction or otherwise.
GP, as projected by the numerical simulations, fared much poorer, retrieving highly varying spectra, cf. Fig.~\ref{fig:noise experiment}(i) and (j). The large standard deviations, especially for the spectral intensity, is caused by the aforementioned tendency of GP to get stuck at local minima. The local minima might share a similar temporal pulse profile---as indicated by the lower deviation of the corresponding temporal intensities---or a FROG trace while having strikingly differing spectra. This behavior is a likely consequence of the weaker features of the SHG FROG trace being obscured by the increasing measurement noise, and the reduced dynamic range due to the extraction of the SHG FROG trace. Local lockups, although a possibility even for evolutionary algorithms, were never observed for DE-iFROG during our tests. As a further merit for DE-iFROG, it should be noted that noise-filtering techniques had to be applied for the extracted SHG FROG traces to improve the results of the GP retrieval, while no effort was made for DE-iFROG to mitigate the noise.

\section{Conclusion}\label{sec:conclusion}

In this work, we have presented a new pulse retrieval algorithm for interferometric frequency-resolved optical gating measurements, named as DE-iFROG. The DE-iFROG is based on the genetic algorithm \emph{differential evolution}, and can use all the available subtraces contained in an iFROG measurement simultaneously. Pulse retrieval is possible even for an incomplete data set or a severely impaired measurement.

The DE-iFROG algorithm was tested using data sets simulating a score of possible errors that may be encountered in real-life measurements. The test cases include additive noise, imperfect spectral response due to a finite phase matching bandwidth, and timing jitter of the delay line. Even though the numerically-introduced errors were massive in comparison to realistic scenarios for modern optics laboratories, the DE-iFROG algorithm reconstructed both the intensity and phase of the highly structured test pulse effortlessly. Through numerical simulations, the noise resilience and accuracy of DE-iFROG pulse retrieval was proven superior to a commonly used procedure, where a SHG FROG trace is first extracted from a SHG iFROG measurement, and then fed into a generalized-projections-based FROG algorithm.

In addition to the comprehensive simulations, the DE-iFROG method was demonstrated experimentally using unamplified near-infrared few-cycle pulses. The results were in excellent agreement with reference measurements conducted with a commercial SPIDER apparatus. Furthermore, the few-cycle pulses were subsequently subjected to additional dispersion, the amount of which was accurately measured by DE-iFROG. Tests with a highly attenuated iFROG signal before detection were made to further gauge the noise resilience of DE-iFROG. The results strongly corroborated the findings of the numerical simulations, with DE-iFROG largely unaffected by the substantial detection noise. Impressively, DE-iFROG retrieved near-perfect fundamental spectra without any additional information besides the iFROG trace itself. On the contrary, GP suffered from considerably increased uncertainty of the retrieved pulse envelope. This behavior was attributed to the increased occurrence of lockups, where the algorithm falters at a local minimum.

In light of the presented results, we believe our method can make iFROG a viable alternative to the established pulse characterization techniques, especially for few-cycle pulses. The collinear beam geometry allows tight focusing and avoids geometrical beam smearing, while simplifying the alignment and reducing the footprint of the FROG setup.
The immense noise resilience of DE-iFROG makes the method a highly valuable tool for situations where only a limited signal-to-noise ratio can be achieved. Yet another, further specialized use is \textit{in situ} pulse characterization where the nonlinear medium itself is studied. Furthermore, the temporal characteristics in a focus of a high numerical aperture objective\cite{fittinghoff_collinear_1998}, or a nanoscale light source can be investigated\cite{berweger_femtosecond_2011,schmidt_adiabatic_2012}.

While the present work focused on adapting differential evolution algorithm for second-harmonic generation iFROG, the algorithm can readily be adapted for iFROG measurements employing third-harmonic generation instead. Furthermore, adaptation for other FROG-like pulse characterization techniques, such as dispersion scan\cite{miranda_simultaneous_2012} is possible. The possibility of using a new type of algorithm for these established measurement techniques is especially intriguing, as the relatively complicated analytical models describing the measurements prohibit the use of generalized projections, which is normally considered the weapon of choice for pulse shape retrieval.


%
%

%


\bibliography{DE_IFROG_RSI}

\begin{thebibliography}{29}%
\makeatletter
\providecommand \@ifxundefined [1]{%
 \@ifx{#1\undefined}
}%
\providecommand \@ifnum [1]{%
 \ifnum #1\expandafter \@firstoftwo
 \else \expandafter \@secondoftwo
 \fi
}%
\providecommand \@ifx [1]{%
 \ifx #1\expandafter \@firstoftwo
 \else \expandafter \@secondoftwo
 \fi
}%
\providecommand \natexlab [1]{#1}%
\providecommand \enquote  [1]{``#1''}%
\providecommand \bibnamefont  [1]{#1}%
\providecommand \bibfnamefont [1]{#1}%
\providecommand \citenamefont [1]{#1}%
\providecommand \href@noop [0]{\@secondoftwo}%
\providecommand \href [0]{\begingroup \@sanitize@url \@href}%
\providecommand \@href[1]{\@@startlink{#1}\@@href}%
\providecommand \@@href[1]{\endgroup#1\@@endlink}%
\providecommand \@sanitize@url [0]{\catcode `\\12\catcode `\$12\catcode
  `\&12\catcode `\#12\catcode `\^12\catcode `\_12\catcode `\%12\relax}%
\providecommand \@@startlink[1]{}%
\providecommand \@@endlink[0]{}%
\providecommand \url  [0]{\begingroup\@sanitize@url \@url }%
\providecommand \@url [1]{\endgroup\@href {#1}{\urlprefix }}%
\providecommand \urlprefix  [0]{URL }%
\providecommand \Eprint [0]{\href }%
\providecommand \doibase [0]{http://dx.doi.org/}%
\providecommand \selectlanguage [0]{\@gobble}%
\providecommand \bibinfo  [0]{\@secondoftwo}%
\providecommand \bibfield  [0]{\@secondoftwo}%
\providecommand \translation [1]{[#1]}%
\providecommand \BibitemOpen [0]{}%
\providecommand \bibitemStop [0]{}%
\providecommand \bibitemNoStop [0]{.\EOS\space}%
\providecommand \EOS [0]{\spacefactor3000\relax}%
\providecommand \BibitemShut  [1]{\csname bibitem#1\endcsname}%
\let\auto@bib@innerbib\@empty
\bibitem [{\citenamefont {Trebino}(2000)}]{trebino_book_2000}%
  \BibitemOpen
  \bibfield  {author} {\bibinfo {author} {\bibfnamefont {R.}~\bibnamefont
  {Trebino}},\ }\href@noop {} {\emph {\bibinfo {title} {{Frequency-Resolved
  Optical Gating: The Measurement of Ultrashort Laser Pulses}}}}\ (\bibinfo
  {publisher} {Springer},\ \bibinfo {year} {2000})\BibitemShut {NoStop}%
\bibitem [{\citenamefont {Baltu\v{s}ka}, \citenamefont {Pshenichnikov},\ and\
  \citenamefont {Wiersma}(1998)}]{baltuska_amplitude_1998}%
  \BibitemOpen
  \bibfield  {author} {\bibinfo {author} {\bibfnamefont {A.}~\bibnamefont
  {Baltu\v{s}ka}}, \bibinfo {author} {\bibfnamefont {M.~S.}\ \bibnamefont
  {Pshenichnikov}}, \ and\ \bibinfo {author} {\bibfnamefont {D.~A.}\
  \bibnamefont {Wiersma}},\ }\href {\doibase 10.1364/OL.23.001474} {\bibfield
  {journal} {\bibinfo  {journal} {Opt. Lett.}\ }\textbf {\bibinfo {volume}
  {23}},\ \bibinfo {pages} {1474} (\bibinfo {year} {1998})}\BibitemShut
  {NoStop}%
\bibitem [{\citenamefont {Fittinghoff}\ \emph {et~al.}(1998)\citenamefont
  {Fittinghoff}, \citenamefont {Squier}, \citenamefont {Barty}, \citenamefont
  {Sweetser}, \citenamefont {Trebino},\ and\ \citenamefont
  {M\"{u}ller}}]{fittinghoff_collinear_1998}%
  \BibitemOpen
  \bibfield  {author} {\bibinfo {author} {\bibfnamefont {D.~N.}\ \bibnamefont
  {Fittinghoff}}, \bibinfo {author} {\bibfnamefont {J.~A.}\ \bibnamefont
  {Squier}}, \bibinfo {author} {\bibfnamefont {C.~P.~J.}\ \bibnamefont
  {Barty}}, \bibinfo {author} {\bibfnamefont {J.~N.}\ \bibnamefont {Sweetser}},
  \bibinfo {author} {\bibfnamefont {R.}~\bibnamefont {Trebino}}, \ and\
  \bibinfo {author} {\bibfnamefont {M.}~\bibnamefont {M\"{u}ller}},\ }\href
  {\doibase 10.1364/OL.23.001046} {\bibfield  {journal} {\bibinfo  {journal}
  {Opt. Lett.}\ }\textbf {\bibinfo {volume} {23}},\ \bibinfo {pages} {1046}
  (\bibinfo {year} {1998})}\BibitemShut {NoStop}%
\bibitem [{\citenamefont {Amat-Rold\'{a}n}\ \emph {et~al.}(2004)\citenamefont
  {Amat-Rold\'{a}n}, \citenamefont {Cormack}, \citenamefont {Loza-Alvarez},
  \citenamefont {Gualda},\ and\ \citenamefont
  {Artigas}}]{amat-roldan_ultrashort_2004}%
  \BibitemOpen
  \bibfield  {author} {\bibinfo {author} {\bibfnamefont {I.}~\bibnamefont
  {Amat-Rold\'{a}n}}, \bibinfo {author} {\bibfnamefont {I.~G.}\ \bibnamefont
  {Cormack}}, \bibinfo {author} {\bibfnamefont {P.}~\bibnamefont
  {Loza-Alvarez}}, \bibinfo {author} {\bibfnamefont {E.~J.}\ \bibnamefont
  {Gualda}}, \ and\ \bibinfo {author} {\bibfnamefont {D.}~\bibnamefont
  {Artigas}},\ }\href {\doibase 10.1364/OPEX.12.001169} {\bibfield  {journal}
  {\bibinfo  {journal} {Opt. Express}\ }\textbf {\bibinfo {volume} {12}},\
  \bibinfo {pages} {1169} (\bibinfo {year} {2004})}\BibitemShut {NoStop}%
\bibitem [{\citenamefont {Stibenz}\ and\ \citenamefont
  {Steinmeyer}(2006{\natexlab{a}})}]{stibenz_structures_2006}%
  \BibitemOpen
  \bibfield  {author} {\bibinfo {author} {\bibfnamefont {G.}~\bibnamefont
  {Stibenz}}\ and\ \bibinfo {author} {\bibfnamefont {G.}~\bibnamefont
  {Steinmeyer}},\ }\href@noop {} {\bibfield  {journal} {\bibinfo  {journal}
  {IEEE Journal of selected topics in quantum electronics}\ }\textbf {\bibinfo
  {volume} {12}},\ \bibinfo {pages} {286} (\bibinfo {year}
  {2006}{\natexlab{a}})}\BibitemShut {NoStop}%
\bibitem [{\citenamefont {Berweger}\ \emph {et~al.}(2011)\citenamefont
  {Berweger}, \citenamefont {Atkin}, \citenamefont {Xu}, \citenamefont
  {Olmon},\ and\ \citenamefont {Raschke}}]{berweger_femtosecond_2011}%
  \BibitemOpen
  \bibfield  {author} {\bibinfo {author} {\bibfnamefont {S.}~\bibnamefont
  {Berweger}}, \bibinfo {author} {\bibfnamefont {J.~M.}\ \bibnamefont {Atkin}},
  \bibinfo {author} {\bibfnamefont {X.~G.}\ \bibnamefont {Xu}}, \bibinfo
  {author} {\bibfnamefont {R.~L.}\ \bibnamefont {Olmon}}, \ and\ \bibinfo
  {author} {\bibfnamefont {M.~B.}\ \bibnamefont {Raschke}},\ }\href@noop {}
  {\bibfield  {journal} {\bibinfo  {journal} {Nano letters}\ }\textbf {\bibinfo
  {volume} {11}},\ \bibinfo {pages} {4309} (\bibinfo {year}
  {2011})}\BibitemShut {NoStop}%
\bibitem [{\citenamefont {Schmidt}\ \emph {et~al.}(2012)\citenamefont
  {Schmidt}, \citenamefont {Piglosiewicz}, \citenamefont {Sadiq}, \citenamefont
  {Shirdel}, \citenamefont {Lee}, \citenamefont {Vasa}, \citenamefont {Park},
  \citenamefont {Kim},\ and\ \citenamefont {Lienau}}]{schmidt_adiabatic_2012}%
  \BibitemOpen
  \bibfield  {author} {\bibinfo {author} {\bibfnamefont {S.}~\bibnamefont
  {Schmidt}}, \bibinfo {author} {\bibfnamefont {B.}~\bibnamefont
  {Piglosiewicz}}, \bibinfo {author} {\bibfnamefont {D.}~\bibnamefont {Sadiq}},
  \bibinfo {author} {\bibfnamefont {J.}~\bibnamefont {Shirdel}}, \bibinfo
  {author} {\bibfnamefont {J.~S.}\ \bibnamefont {Lee}}, \bibinfo {author}
  {\bibfnamefont {P.}~\bibnamefont {Vasa}}, \bibinfo {author} {\bibfnamefont
  {N.}~\bibnamefont {Park}}, \bibinfo {author} {\bibfnamefont {D.-S.}\
  \bibnamefont {Kim}}, \ and\ \bibinfo {author} {\bibfnamefont
  {C.}~\bibnamefont {Lienau}},\ }\href@noop {} {\bibfield  {journal} {\bibinfo
  {journal} {ACS nano}\ }\textbf {\bibinfo {volume} {6}},\ \bibinfo {pages}
  {6040} (\bibinfo {year} {2012})}\BibitemShut {NoStop}%
\bibitem [{\citenamefont {Hofmann}\ \emph {et~al.}(2015)\citenamefont
  {Hofmann}, \citenamefont {Hyyti}, \citenamefont {Birkholz}, \citenamefont
  {Bock}, \citenamefont {Das}, \citenamefont {Grunwald}, \citenamefont
  {Hoffmann}, \citenamefont {Nagy}, \citenamefont {Demircan}, \citenamefont
  {Jup\'{e}}, \citenamefont {Ristau}, \citenamefont {Morgner}, \citenamefont
  {Br\'{e}e}, \citenamefont {Woerner}, \citenamefont {Elsaesser},\ and\
  \citenamefont {Steinmeyer}}]{hofmann_noninstantaneous_2015}%
  \BibitemOpen
  \bibfield  {author} {\bibinfo {author} {\bibfnamefont {M.}~\bibnamefont
  {Hofmann}}, \bibinfo {author} {\bibfnamefont {J.}~\bibnamefont {Hyyti}},
  \bibinfo {author} {\bibfnamefont {S.}~\bibnamefont {Birkholz}}, \bibinfo
  {author} {\bibfnamefont {M.}~\bibnamefont {Bock}}, \bibinfo {author}
  {\bibfnamefont {S.~K.}\ \bibnamefont {Das}}, \bibinfo {author} {\bibfnamefont
  {R.}~\bibnamefont {Grunwald}}, \bibinfo {author} {\bibfnamefont
  {M.}~\bibnamefont {Hoffmann}}, \bibinfo {author} {\bibfnamefont
  {T.}~\bibnamefont {Nagy}}, \bibinfo {author} {\bibfnamefont {A.}~\bibnamefont
  {Demircan}}, \bibinfo {author} {\bibfnamefont {M.}~\bibnamefont {Jup\'{e}}},
  \bibinfo {author} {\bibfnamefont {D.}~\bibnamefont {Ristau}}, \bibinfo
  {author} {\bibfnamefont {U.}~\bibnamefont {Morgner}}, \bibinfo {author}
  {\bibfnamefont {C.}~\bibnamefont {Br\'{e}e}}, \bibinfo {author}
  {\bibfnamefont {M.}~\bibnamefont {Woerner}}, \bibinfo {author} {\bibfnamefont
  {T.}~\bibnamefont {Elsaesser}}, \ and\ \bibinfo {author} {\bibfnamefont
  {G.}~\bibnamefont {Steinmeyer}},\ }\href {\doibase 10.1364/OPTICA.2.000151}
  {\bibfield  {journal} {\bibinfo  {journal} {Optica}\ }\textbf {\bibinfo
  {volume} {2}},\ \bibinfo {pages} {151} (\bibinfo {year} {2015})}\BibitemShut
  {NoStop}%
\bibitem [{\citenamefont {Metzger}, \citenamefont {Steinmann},\ and\
  \citenamefont {Giessen}(2011)}]{metzger_high-power_2011}%
  \BibitemOpen
  \bibfield  {author} {\bibinfo {author} {\bibfnamefont {B.}~\bibnamefont
  {Metzger}}, \bibinfo {author} {\bibfnamefont {A.}~\bibnamefont {Steinmann}},
  \ and\ \bibinfo {author} {\bibfnamefont {H.}~\bibnamefont {Giessen}},\
  }\href@noop {} {\bibfield  {journal} {\bibinfo  {journal} {Optics express}\
  }\textbf {\bibinfo {volume} {19}},\ \bibinfo {pages} {24354} (\bibinfo {year}
  {2011})}\BibitemShut {NoStop}%
\bibitem [{\citenamefont {Stibenz}\ and\ \citenamefont
  {Steinmeyer}(2005)}]{stibenz_interferometric_2005}%
  \BibitemOpen
  \bibfield  {author} {\bibinfo {author} {\bibfnamefont {G.}~\bibnamefont
  {Stibenz}}\ and\ \bibinfo {author} {\bibfnamefont {G.}~\bibnamefont
  {Steinmeyer}},\ }\href
  {http://www.opticsinfobase.org/abstract.cfm?uri=OE-13-7-2617} {\bibfield
  {journal} {\bibinfo  {journal} {Opt. Express}\ }\textbf {\bibinfo {volume}
  {13}},\ \bibinfo {pages} {2617} (\bibinfo {year} {2005})}\BibitemShut
  {NoStop}%
\bibitem [{\citenamefont {Storn}\ and\ \citenamefont
  {Price}(1997)}]{storn_differential_1997}%
  \BibitemOpen
  \bibfield  {author} {\bibinfo {author} {\bibfnamefont {R.}~\bibnamefont
  {Storn}}\ and\ \bibinfo {author} {\bibfnamefont {K.}~\bibnamefont {Price}},\
  }\href@noop {} {\bibfield  {journal} {\bibinfo  {journal} {Journal of global
  optimization}\ }\textbf {\bibinfo {volume} {11}},\ \bibinfo {pages} {341}
  (\bibinfo {year} {1997})}\BibitemShut {NoStop}%
\bibitem [{\citenamefont {Stibenz}\ and\ \citenamefont
  {Steinmeyer}(2006{\natexlab{b}})}]{stibenz_optimizing_2006}%
  \BibitemOpen
  \bibfield  {author} {\bibinfo {author} {\bibfnamefont {G.}~\bibnamefont
  {Stibenz}}\ and\ \bibinfo {author} {\bibfnamefont {G.}~\bibnamefont
  {Steinmeyer}},\ }\href {\doibase 10.1063/1.2221511} {\bibfield  {journal}
  {\bibinfo  {journal} {Review of Scientific Instruments}\ }\textbf {\bibinfo
  {volume} {77}},\ \bibinfo {pages} {073105} (\bibinfo {year}
  {2006}{\natexlab{b}})},\ \Eprint
  {http://arxiv.org/abs/http://dx.doi.org/10.1063/1.2221511}
  {http://dx.doi.org/10.1063/1.2221511} \BibitemShut {NoStop}%
\bibitem [{\citenamefont {Sala}, \citenamefont {Kenney-Wallace},\ and\
  \citenamefont {Hall}(1980)}]{sala_correlation_1980}%
  \BibitemOpen
  \bibfield  {author} {\bibinfo {author} {\bibfnamefont {K.}~\bibnamefont
  {Sala}}, \bibinfo {author} {\bibfnamefont {G.}~\bibnamefont
  {Kenney-Wallace}}, \ and\ \bibinfo {author} {\bibfnamefont {G.}~\bibnamefont
  {Hall}},\ }\href {\doibase 10.1109/JQE.1980.1070606} {\bibfield  {journal}
  {\bibinfo  {journal} {IEEE Journal of Quantum Electronics}\ }\textbf
  {\bibinfo {volume} {16}},\ \bibinfo {pages} {990} (\bibinfo {year}
  {1980})}\BibitemShut {NoStop}%
\bibitem [{\citenamefont {Diels}\ \emph {et~al.}(1985)\citenamefont {Diels},
  \citenamefont {Fontaine}, \citenamefont {McMichael},\ and\ \citenamefont
  {Simoni}}]{diels_control_1985}%
  \BibitemOpen
  \bibfield  {author} {\bibinfo {author} {\bibfnamefont {J.-C.~M.}\
  \bibnamefont {Diels}}, \bibinfo {author} {\bibfnamefont {J.~J.}\ \bibnamefont
  {Fontaine}}, \bibinfo {author} {\bibfnamefont {I.~C.}\ \bibnamefont
  {McMichael}}, \ and\ \bibinfo {author} {\bibfnamefont {F.}~\bibnamefont
  {Simoni}},\ }\href {\doibase 10.1364/AO.24.001270} {\bibfield  {journal}
  {\bibinfo  {journal} {Appl. Opt.}\ }\textbf {\bibinfo {volume} {24}},\
  \bibinfo {pages} {1270} (\bibinfo {year} {1985})}\BibitemShut {NoStop}%
\bibitem [{\citenamefont {Hirayama}\ and\ \citenamefont
  {Sheik-Bahae}(2002)}]{hirayama_realtime_2002}%
  \BibitemOpen
  \bibfield  {author} {\bibinfo {author} {\bibfnamefont {T.}~\bibnamefont
  {Hirayama}}\ and\ \bibinfo {author} {\bibfnamefont {M.}~\bibnamefont
  {Sheik-Bahae}},\ }\href {\doibase 10.1364/OL.27.000860} {\bibfield  {journal}
  {\bibinfo  {journal} {Opt. Lett.}\ }\textbf {\bibinfo {volume} {27}},\
  \bibinfo {pages} {860} (\bibinfo {year} {2002})}\BibitemShut {NoStop}%
\bibitem [{\citenamefont {Nicholson}\ and\ \citenamefont
  {Rudolph}(2002)}]{nicholson_noise_2002}%
  \BibitemOpen
  \bibfield  {author} {\bibinfo {author} {\bibfnamefont {J.~W.}\ \bibnamefont
  {Nicholson}}\ and\ \bibinfo {author} {\bibfnamefont {W.}~\bibnamefont
  {Rudolph}},\ }\href {\doibase 10.1364/JOSAB.19.000330} {\bibfield  {journal}
  {\bibinfo  {journal} {J. Opt. Soc. Am. B}\ }\textbf {\bibinfo {volume}
  {19}},\ \bibinfo {pages} {330} (\bibinfo {year} {2002})}\BibitemShut
  {NoStop}%
\bibitem [{\citenamefont {DeLong}\ \emph {et~al.}(1994)\citenamefont {DeLong},
  \citenamefont {Kohler}, \citenamefont {Wilson}, \citenamefont {Fittinghoff},\
  and\ \citenamefont {Trebino}}]{delong_pulse_1994}%
  \BibitemOpen
  \bibfield  {author} {\bibinfo {author} {\bibfnamefont {K.~W.}\ \bibnamefont
  {DeLong}}, \bibinfo {author} {\bibfnamefont {B.}~\bibnamefont {Kohler}},
  \bibinfo {author} {\bibfnamefont {K.}~\bibnamefont {Wilson}}, \bibinfo
  {author} {\bibfnamefont {D.~N.}\ \bibnamefont {Fittinghoff}}, \ and\ \bibinfo
  {author} {\bibfnamefont {R.}~\bibnamefont {Trebino}},\ }\href {\doibase
  10.1364/OL.19.002152} {\bibfield  {journal} {\bibinfo  {journal} {Opt.
  Lett.}\ }\textbf {\bibinfo {volume} {19}},\ \bibinfo {pages} {2152} (\bibinfo
  {year} {1994})}\BibitemShut {NoStop}%
\bibitem [{\citenamefont {Rhodes}\ \emph {et~al.}(2013)\citenamefont {Rhodes},
  \citenamefont {Steinmeyer}, \citenamefont {Ratner},\ and\ \citenamefont
  {Trebino}}]{rhodes_pulseshape_2013}%
  \BibitemOpen
  \bibfield  {author} {\bibinfo {author} {\bibfnamefont {M.}~\bibnamefont
  {Rhodes}}, \bibinfo {author} {\bibfnamefont {G.}~\bibnamefont {Steinmeyer}},
  \bibinfo {author} {\bibfnamefont {J.}~\bibnamefont {Ratner}}, \ and\ \bibinfo
  {author} {\bibfnamefont {R.}~\bibnamefont {Trebino}},\ }\href {\doibase
  10.1002/lpor.201200102} {\bibfield  {journal} {\bibinfo  {journal} {Laser \&
  Photonics Reviews}\ }\textbf {\bibinfo {volume} {7}},\ \bibinfo {pages} {557}
  (\bibinfo {year} {2013})}\BibitemShut {NoStop}%
\bibitem [{\citenamefont {Amat-Rold\'{a}n}\ \emph {et~al.}(2005)\citenamefont
  {Amat-Rold\'{a}n}, \citenamefont {Cormack}, \citenamefont {Loza-Alvarez},\
  and\ \citenamefont {Artigas}}]{amat-roldan_measurement_2005}%
  \BibitemOpen
  \bibfield  {author} {\bibinfo {author} {\bibfnamefont {I.}~\bibnamefont
  {Amat-Rold\'{a}n}}, \bibinfo {author} {\bibfnamefont {I.~G.}\ \bibnamefont
  {Cormack}}, \bibinfo {author} {\bibfnamefont {P.}~\bibnamefont
  {Loza-Alvarez}}, \ and\ \bibinfo {author} {\bibfnamefont {D.}~\bibnamefont
  {Artigas}},\ }\href {\doibase 10.1364/OL.30.001063} {\bibfield  {journal}
  {\bibinfo  {journal} {Opt. Lett.}\ }\textbf {\bibinfo {volume} {30}},\
  \bibinfo {pages} {1063} (\bibinfo {year} {2005})}\BibitemShut {NoStop}%
\bibitem [{\citenamefont {Wu}\ \emph {et~al.}(2016)\citenamefont {Wu},
  \citenamefont {Lu}, \citenamefont {Weng}, \citenamefont {Chen},\ and\
  \citenamefont {Yang}}]{wu_dispersion_2016}%
  \BibitemOpen
  \bibfield  {author} {\bibinfo {author} {\bibfnamefont {P.-Y.}\ \bibnamefont
  {Wu}}, \bibinfo {author} {\bibfnamefont {H.-H.}\ \bibnamefont {Lu}}, \bibinfo
  {author} {\bibfnamefont {C.-Z.}\ \bibnamefont {Weng}}, \bibinfo {author}
  {\bibfnamefont {Y.-H.}\ \bibnamefont {Chen}}, \ and\ \bibinfo {author}
  {\bibfnamefont {S.-D.}\ \bibnamefont {Yang}},\ }\href {\doibase
  10.1364/OL.41.004538} {\bibfield  {journal} {\bibinfo  {journal} {Opt.
  Lett.}\ }\textbf {\bibinfo {volume} {41}},\ \bibinfo {pages} {4538} (\bibinfo
  {year} {2016})}\BibitemShut {NoStop}%
\bibitem [{\citenamefont {Nicholson}\ \emph {et~al.}(1999)\citenamefont
  {Nicholson}, \citenamefont {Omenetto}, \citenamefont {Funk},\ and\
  \citenamefont {Taylor}}]{nicholson_evolving_1999}%
  \BibitemOpen
  \bibfield  {author} {\bibinfo {author} {\bibfnamefont {J.~W.}\ \bibnamefont
  {Nicholson}}, \bibinfo {author} {\bibfnamefont {F.~G.}\ \bibnamefont
  {Omenetto}}, \bibinfo {author} {\bibfnamefont {D.~J.}\ \bibnamefont {Funk}},
  \ and\ \bibinfo {author} {\bibfnamefont {A.~J.}\ \bibnamefont {Taylor}},\
  }\href {\doibase 10.1364/OL.24.000490} {\bibfield  {journal} {\bibinfo
  {journal} {Opt. Lett.}\ }\textbf {\bibinfo {volume} {24}},\ \bibinfo {pages}
  {490} (\bibinfo {year} {1999})}\BibitemShut {NoStop}%
\bibitem [{\citenamefont {Rocca}, \citenamefont {Oliveri},\ and\ \citenamefont
  {Massa}(2011)}]{rocca_differential_2011}%
  \BibitemOpen
  \bibfield  {author} {\bibinfo {author} {\bibfnamefont {P.}~\bibnamefont
  {Rocca}}, \bibinfo {author} {\bibfnamefont {G.}~\bibnamefont {Oliveri}}, \
  and\ \bibinfo {author} {\bibfnamefont {A.}~\bibnamefont {Massa}},\ }\href
  {\doibase 10.1109/MAP.2011.5773566} {\bibfield  {journal} {\bibinfo
  {journal} {IEEE Antennas and Propagation Magazine}\ }\textbf {\bibinfo
  {volume} {53}},\ \bibinfo {pages} {38} (\bibinfo {year} {2011})}\BibitemShut
  {NoStop}%
\bibitem [{\citenamefont {Sidorenko}\ \emph {et~al.}(2016)\citenamefont
  {Sidorenko}, \citenamefont {Lahav}, \citenamefont {Avnat},\ and\
  \citenamefont {Cohen}}]{sidorenko_ptychographic_2016}%
  \BibitemOpen
  \bibfield  {author} {\bibinfo {author} {\bibfnamefont {P.}~\bibnamefont
  {Sidorenko}}, \bibinfo {author} {\bibfnamefont {O.}~\bibnamefont {Lahav}},
  \bibinfo {author} {\bibfnamefont {Z.}~\bibnamefont {Avnat}}, \ and\ \bibinfo
  {author} {\bibfnamefont {O.}~\bibnamefont {Cohen}},\ }\href {\doibase
  10.1364/OPTICA.3.001320} {\bibfield  {journal} {\bibinfo  {journal} {Optica}\
  }\textbf {\bibinfo {volume} {3}},\ \bibinfo {pages} {1320} (\bibinfo {year}
  {2016})}\BibitemShut {NoStop}%
\bibitem [{\citenamefont {Hyyti}\ \emph {et~al.}(2017)\citenamefont {Hyyti},
  \citenamefont {Escoto}, \citenamefont {Steinmeyer},\ and\ \citenamefont
  {Witting}}]{hyyti_interferometric_2017}%
  \BibitemOpen
  \bibfield  {author} {\bibinfo {author} {\bibfnamefont {J.}~\bibnamefont
  {Hyyti}}, \bibinfo {author} {\bibfnamefont {E.}~\bibnamefont {Escoto}},
  \bibinfo {author} {\bibfnamefont {G.}~\bibnamefont {Steinmeyer}}, \ and\
  \bibinfo {author} {\bibfnamefont {T.}~\bibnamefont {Witting}},\ }\href@noop
  {} {\bibfield  {journal} {\bibinfo  {journal} {Optics Letters}\ }\textbf
  {\bibinfo {volume} {42}},\ \bibinfo {pages} {2185} (\bibinfo {year}
  {2017})}\BibitemShut {NoStop}%
\bibitem [{\citenamefont {Kane}\ \emph {et~al.}(1997)\citenamefont {Kane},
  \citenamefont {Rodriguez}, \citenamefont {Taylor},\ and\ \citenamefont
  {Clement}}]{kane_simultaneous_1997}%
  \BibitemOpen
  \bibfield  {author} {\bibinfo {author} {\bibfnamefont {D.~J.}\ \bibnamefont
  {Kane}}, \bibinfo {author} {\bibfnamefont {G.}~\bibnamefont {Rodriguez}},
  \bibinfo {author} {\bibfnamefont {A.~J.}\ \bibnamefont {Taylor}}, \ and\
  \bibinfo {author} {\bibfnamefont {T.~S.}\ \bibnamefont {Clement}},\ }\href
  {\doibase 10.1364/JOSAB.14.000935} {\bibfield  {journal} {\bibinfo  {journal}
  {J. Opt. Soc. Am. B}\ }\textbf {\bibinfo {volume} {14}},\ \bibinfo {pages}
  {935} (\bibinfo {year} {1997})}\BibitemShut {NoStop}%
\bibitem [{\citenamefont {Fittinghoff}\ \emph {et~al.}(1995)\citenamefont
  {Fittinghoff}, \citenamefont {DeLong}, \citenamefont {Trebino},\ and\
  \citenamefont {Ladera}}]{fittinghoff_noise_1995}%
  \BibitemOpen
  \bibfield  {author} {\bibinfo {author} {\bibfnamefont {D.~N.}\ \bibnamefont
  {Fittinghoff}}, \bibinfo {author} {\bibfnamefont {K.~W.}\ \bibnamefont
  {DeLong}}, \bibinfo {author} {\bibfnamefont {R.}~\bibnamefont {Trebino}}, \
  and\ \bibinfo {author} {\bibfnamefont {C.~L.}\ \bibnamefont {Ladera}},\
  }\href {\doibase 10.1364/JOSAB.12.001955} {\bibfield  {journal} {\bibinfo
  {journal} {J. Opt. Soc. Am. B}\ }\textbf {\bibinfo {volume} {12}},\ \bibinfo
  {pages} {1955} (\bibinfo {year} {1995})}\BibitemShut {NoStop}%
\bibitem [{fro()}]{frog_code}%
  \BibitemOpen
  \href@noop {} {}\bibinfo {note} {FROG code version 1.2, Georgia Tech, Trebino
  group (2012). Available at http://frog.gatech.edu/code.html}\BibitemShut
  {NoStop}%
\bibitem [{\citenamefont {Takeda}, \citenamefont {Ina},\ and\ \citenamefont
  {Kobayashi}(1982)}]{takeda_fourier-transform_1982}%
  \BibitemOpen
  \bibfield  {author} {\bibinfo {author} {\bibfnamefont {M.}~\bibnamefont
  {Takeda}}, \bibinfo {author} {\bibfnamefont {H.}~\bibnamefont {Ina}}, \ and\
  \bibinfo {author} {\bibfnamefont {S.}~\bibnamefont {Kobayashi}},\ }\href
  {\doibase 10.1364/JOSA.72.000156} {\bibfield  {journal} {\bibinfo  {journal}
  {J. Opt. Soc. Am.}\ }\textbf {\bibinfo {volume} {72}},\ \bibinfo {pages}
  {156} (\bibinfo {year} {1982})}\BibitemShut {NoStop}%
\bibitem [{\citenamefont {Miranda}\ \emph {et~al.}(2012)\citenamefont
  {Miranda}, \citenamefont {Fordell}, \citenamefont {Arnold}, \citenamefont
  {L'Huillier},\ and\ \citenamefont {Crespo}}]{miranda_simultaneous_2012}%
  \BibitemOpen
  \bibfield  {author} {\bibinfo {author} {\bibfnamefont {M.}~\bibnamefont
  {Miranda}}, \bibinfo {author} {\bibfnamefont {T.}~\bibnamefont {Fordell}},
  \bibinfo {author} {\bibfnamefont {C.}~\bibnamefont {Arnold}}, \bibinfo
  {author} {\bibfnamefont {A.}~\bibnamefont {L'Huillier}}, \ and\ \bibinfo
  {author} {\bibfnamefont {H.}~\bibnamefont {Crespo}},\ }\href {\doibase
  10.1364/OE.20.000688} {\bibfield  {journal} {\bibinfo  {journal} {Opt.
  Express}\ }\textbf {\bibinfo {volume} {20}},\ \bibinfo {pages} {688}
  (\bibinfo {year} {2012})}\BibitemShut {NoStop}%
\end{thebibliography}%


%

\end{document}